\documentclass[fleqn,12pt]{wlscirep}

\usepackage{amsmath,amssymb}
\usepackage{mathrsfs}
\usepackage{listings}

\renewcommand{\d}[1]{\ensuremath{\operatorname{d}\!{#1}}}
\newcommand{\D}[1]{\ensuremath{\operatorname{D}\!{#1}}}
\DeclareMathOperator{\Id}{Id}%
\DeclareMathOperator{\prob}{\mathrm{Pr}}%
\DeclareMathOperator{\Exp}{\mathrm{Ex}}%
\DeclareMathOperator{\area}{area}
\DeclareMathOperator{\vol}{vol}

\def\pct{\%}

\title{Lagrangian stability of the Malvinas Current}

\author[1,*]{F.\ J.\ Beron-Vera}
\author[2,3,4]{N.\ Bodnariuk}
\author[2,3,4]{M.\ Saraceno}
\author[5]{M.\ J.\ Olascoaga}
\author[2,3,4]{C.\ Simionato}

\affil[1]{Department of Atmospheric Sciences, Rosenstiel School of
Marine and Atmospheric Science, University of Miami, Miami, Florida,
USA.} 

\affil[2]{Centro de Investigaciones del Mar y la Atmósfera
(CIMA/CONICET--UBA), Ciudad Aut\'onoma de Buenos Aires,
Argentina.} 

\affil[3]{Departamento de Ciencias de la Atm\'osfera y los Oc\'eanos,
Facultad de Ciencias Exactas y Naturales, Universidad de Buenos
Aires, Ciudad Aut\'onoma de Buenos Aires, Argentina.}

\affil[3]{Unidad Mixta Internacional--Instituto Franco‐Argentino
para el Estudio del Clima y sus Impactos (UMI--IFAECI/CNRS--CONICET--UBA),
Ciudad Aut\'onoma de Buenos Aires, Argentina.}

\affil[5]{Department of Ocean Sciences, Rosenstiel School of Marine
and Atmospheric Science, University of Miami, Miami, Florida, USA.}

\affil[*]{Corresponding author.  E-mail: fberon@rsmas.miami.edu.}

\date{\today}


\begin{abstract}
  Deterministic and probabilistic tools from nonlinear dynamics are
  used to assess enduring near-surface Lagrangian aspects of the
  Malvinas Current.  The deterministic tools are applied on a
  multi-year record of velocities derived from satellite altimetry
  data, revealing a resilient cross-stream transport barrier.  This
  is composed of shearless-parabolic Lagrangian coherent structures
  (LCS), which, extracted over sliding time windows along the
  multi-year altimetry-derived velocity record, lie in near
  coincidental position.  The probabilistic tools are applied on a
  large collection of historical satellite-tracked drifter trajectories,
  revealing weakly communicating flow regions on either side of the
  altimetry-derived barrier.  Shearless-parabolic LCS are detected
  for the first time from altimetry data, and their significance
  is supported on satellite-derived ocean color data, which reveal
  shapes that quite closely resemble the peculiar V shapes, dubbed
  ``chevrons,'' that have recently confirmed the presence of similar
  LCS in the atmosphere of Jupiter.  Finally, using in-situ velocity
  and hydrographic data, conditions for symmetric stability are
  found to be satisfied, suggesting a duality between Lagrangian
  and Eulerian stability for the Malvinas Current.
\end{abstract}

\begin{document}

\flushbottom
\maketitle

\thispagestyle{empty}


\noindent The Malvinas Current originates as a result of a pronounced
northward turn of the northern edge of the Antarctic Circumpolar
Current past the Drake Passage.  Carrying within a substantial
portion of the upper limb of the Atlantic Meridional Overturning
Circulation \cite{Friocourt-etal-05}, it represents a northward
pathway for nutrient-rich subpolar water, making the western margin
of the Argentine Basin a region of enhanced biological activity
\cite{Longhurst-98} and significant fisheries \cite{Acha-etal-04}.
The Malvinas Current flows northward up to about 38$^{\circ}$S,
where it sharply turns eastward upon meeting the southward-flowing
Brazil Current to form the Brazil--Malvinas Confluence \cite{Matano-93},
a region characterized by high mesoscale variability \cite{Goni-etal-96}.
Lagrangian observations have suggested that the Malvinas Current
is composed of a single barotropic jet extending down to 750-m depth
or more for most of its northward path along the western boundary
\cite{Davis-etal-96} as is constrained by potential vorticity
conservation \cite{Saraceno-etal-04, Piola-etal-10}.  High-resolution
hydrographic data and direct current observations more recently
suggested the presence multiple baroclinic jets in addition to the
main barotropic one \cite{Piola-etal-13}, confirming earlier
inferences made from the analysis of the surface thermal structure
\cite{Franco-etal-08}.

The analysis of the surface thermal structure more specifically
revealed regions of large temperature contrast along cores of high
meridional velocity \cite{Piola-etal-13}.  This finding is consistent
with the expectation that the Malvinas Current should behave as
barrier for cross-stream transport.  This expectation is motivated
by behavior of jetstreams in the lower stratosphere \cite{Rypina-etal-07a,
Beron-etal-10a, Beron-JPCS-11, Beron-etal-12, Olascoaga-etal-13}
and the weather layer of Jupiter \cite{Beron-etal-08a,
Hadjighasem-Haller-16} as well as earlier speculation that western
boundary currents such as the Gulf Stream should behave as transport
barriers \cite{Bower-etal-85} and more recent work that has
characterized zonal ocean currents as cross-stream mixing inhibitors
\cite{Ferrari-Nikurashin-10}, which has been partially verified
by applying heuristic analyses involving satellite altimetry data,
drifter trajectories, and ocean color imagery \cite{Marshall-etal-06,
Beron-JGR-10, Rypina-etal-11b, Huhn-etal-12}.  Our goal in this
paper is to test the above expectation and further assess its
persistence over time.

To achieve our goal we use two types of tools from nonlinear dynamics,
both especially designed to investigate global aspects of Lagrangian
motion.  One set of tools is deterministic, and build on geometric,
observer-independent (or objective) notions of strain and shear.
They target so-called \emph{Lagrangian coherent structures}
(\emph{LCS}) \cite{Haller-Yuan-00} as organizers of the Lagrangian
circulation.  This is done by means of a collection of global
variational principles that constitute the geodesic theory of LCS
\cite{Haller-Beron-12, Farazmand-Haller-13, Beron-etal-13,
Haller-Beron-13, Haller-Beron-14, Farazmand-etal-14, Karrasch-etal-14,
Karrasch-15, Haller-16}.  The deterministic tools are more effective
when the velocity field is known as this can be integrated to
generate the required flow map that needs to be subsequently
differentiated with respect to initial positions.

The other set of tools considered is probabilistic.  These tools
root in ergodic theory and, under appropriate time-homogeneity
assumptions, can unveil from the Lagrangian circulation statistically
weak communicating flow regions that form the basis for the
construction of \emph{Lagrangian geographies} \cite{Froyland-etal-14,
Miron-etal-17, Miron-etal-18a, Olascoaga-etal-18}.  The theoretical
foundation for this is provided by a series of results from the
study of autonomous dynamical systems using probability densities
that have led to the notion of \emph{almost-invariant sets}
\cite{Dellnitz-Junge-97, Dellnitz-Junge-99, Froyland-Dellnitz-03,
Froyland-05, Froyland-Padberg-09}.  Central to this approach is the
Perron--Frobenius (or transfer) operator \cite{Lasota-Mackey-94}
and the transition matrix, its discrete version that defines a
Markov chain \cite{Kemeny-Snell-76, Horn-Johnson-90, Norris-98} on
boxes covering the flow domain.  The probabilistic tools do not
require flow map differentiation and can be applied directly on
Lagrangian trajectories that do not start simultaneously under the
above assumptions.

The deterministic tools are applied on a multi-year record of
velocities derived from satellite altimetry data, revealing a
persisting cross-stream transport barrier associated with the
Malvinas Current in the near surface ocean.  This barrier is composed
of \emph{shearless-parabolic} LCS, which, extracted over sliding
time windows along the multi-year altimetry-derived velocity record,
lie in near-coincidental position.  Shearless-parabolic LCS generalize
the concept of twistless invariant KAM (Kolmogorov--Arnold--Moser)
tori from time-periodic \cite{delCastillo-Morrison-93,
Delshams-delaLlave-00} or quasiperiodic \cite{Rypina-etal-07b,
Beron-etal-10a} flows to finite-time-aperiodic flows.  The probabilistic
tools are applied on a large collection of historical satellite-tracked
trajectories of drifters drogued at 15 m, revealing statistically
weak communicating flow regions on either side of the altimetry-derived
barrier.

Shearless-parabolic LCS are detected for the first time from altimetry
data, and their significance is supported on satellite-derived ocean
color data.  Patterns revealed in such data are found to organize
into V shapes nearly axially straddling the altimetry-derived LCS
consistent with their shearless-parabolic nature \cite{Farazmand-etal-14}.
Such V shapes have been to the best of our knowledge only reported
to develop by clouds in the atmosphere of Jupiter at the boundaries
of its characteristic zonal strips \cite{Simon-etal-12}. The
significance of the enduring cross-stream transport barrier which
characterizes the Malvinas Current is independently supported on
drifter data. Furthermore, in-situ velocity and hydrographic data
are used to suggest a duality between Lagrangian and Eulerian
stability for the current.

The rest of the paper follows the standard organization into a
methods section, a results section, and a summary and conclusions
section. The various types of data considered are described as they
are employed.  Finally, the online Supplementary Information provides
additional details on the deterministic and probabilistic tools as
well as on the Eulerian stability result considered, thereby making
the paper quite selfcontained.

\section*{Methods}

\paragraph{Deterministic tools.}

The deterministic procedure involved in shearless-parabolic LCS
extraction is succinctly as follows  (cf.\ Section A in the online
Supplementary Information for an expanded discussion).  Given an
incompressible, two-dimensional velocity field $v(x,t)$, the fluid
trajectory equation
\begin{equation}
  \dot{x} = v(x,t)
  \label{eq:v}
\end{equation}
is integrated over the time interval $[t_0,t]$ to form the Cauchy--Green
tensor
\begin{equation}
  C_{t_0}^t(x_0) = \D{F}_{t_0}^t(x_0)^\top \D{F}_{t_0}^t(x_0),
  \label{eq:CG}
\end{equation}
where $F_{t_0}^t$ is the flow map that takes initial positions $x_0$
at $t_0$ to positions $x(t;x_0,t_0)$ at $t$. The eigenvalues and
eigenvectors of this observer-independent (i.e., objective) measure
of deformation, satisfying
\begin{equation}
  0 < \lambda_1(x_0)\equiv \lambda_2(x_0)^{-1} \leq 1,\quad
  \xi_1(x_0) \perp \xi_2(x_0),
\end{equation}
are then computed.

Time-$t_0$ positions of \emph{shearless-parabolic LCS} are subsequently
sought as material lines formed by chains of alternating segments
of Cauchy--Green tensorlines everywhere tangent to $\xi_1(x_0)$ and
$\xi_2(x_0)$, namely, curves $r(s)$ satisfying
\begin{equation}
  r' = \xi_i(r),\quad i=1,2.
  \label{eq:xi}
\end{equation}
The $\xi_1(x_0)$- and $\xi_2(x_0)$-line segments are chosen:
\begin{enumerate}
  \item to connect wedge and trisector singularities where
  $C_{t_0}^t(x_0) = \Id$, so the construction is structurally stable
  (i.e., robust under flow map perturbations); and
  \item such that $\smash{\sqrt{\lambda_1(x_0)}} \approx 1$ and
  $\smash{\sqrt{\lambda_2(x_0)}} \approx 1$, respectively, so
  along-segment squeezing and stretching is kept close to neutral,
  thereby minimizing their hyperbolic nature.
\end{enumerate}

A disk filled with tracer which is initially divided into two halves
by one such shearless-parabolic LCS will, therefore, deform into a
V shape axially straddling the LCS \cite{Farazmand-etal-14}.
Observation of such behavior represents a more stringent test of
the presence of a shearless-parabolic LCS with a cross-stream
transport inhibitor effect than the observation of a large tracer
contrast, which can be produced by LCS of any type.

\paragraph{Probabilistic tools.}

Central to the probabilistic approach is a transition matrix $P\in
\mathbb{R}^{N \times N}$ with components  (cf.\ Section B in the
online Supplementary Information for details)
\begin{equation}
  P_{ij} := \prob[\xi_{t+\mathcal T}\in B_j \mid \xi_t \in B_i]
  \label{eq:Pij}
\end{equation}
for a given transition time $\mathcal T$ and any time $t$, which
provides a discrete representation of a transfer operator for the
passive evolution on a domain $X$ of tracers governed by a
time-homogeneous advection--diffusion process.  Drawn from a uniform
distribution on $B_i$, $\xi_t$ is the initial position of a
discrete-time trajectory $\{\xi_t, \xi_{t+\mathcal T}, \xi_{t+2\mathcal
T}, \dotsc\}$ described by a tracer particle that randomly jumps,
according to the stochastic kernel of the advection--diffusion
process, between boxes in a collection $\smash{\{B_1, \dotsc, B_N\}}$
covering $X$.  If $X$ is closed, then $\smash{\sum_{j=1}^N} P_{ij}
= 1$ for every $i$, i.e., $P$ is row-stochastic.  By construction,
$P$ defines a Markov chain with states represented by the boxes of
the partition.  The discrete representation of a tracer probability
density $f(x)$, i.e., satisfying $\smash{\int_X}f(x)\d{x} = 1$, is
a probability vector $\mathbf{f} = (f_1, \cdots ,f_N)$, where $f_i
= \int_{B_i} f(x)\d{x}$ so $\smash{\sum_{i=1}^N} f_i = 1$.  This
evolves $k\mathcal T$ units of time according to
\begin{equation}
  \mathbf f^{(k)} = \mathbf f P^k,\quad k = 1, 2, \dotsc.
\end{equation}
Long-time asymptotic aspects of the evolution on the Markov chain
described $P$ can be inferred from its spectral properties \cite{Hsu-87,
Dellnitz-Junge-99, Froyland-05} as follows.

If $P$ is irreducible (i.e., all states in the Markov chain
communicate) and aperiodic (i.e., no state is revisited cyclically),
its dominant \emph{left} eigenvector, $\mathbf p$, satisfies $\mathbf
pP = \mathbf p$, can be chosen (componentwise) positive, and (scaled
appropriately) represents a \emph{limiting invariant distribution},
namely, $\mathbf{p} = \smash{\lim_{k\uparrow \infty} \mathbf f P^k}$
for any probability vector $\mathbf f$ (cf., e.g., Horn and Johnson
\cite{Horn-Johnson-90}).  In particular, $\mathbf f = \mathbf 1/N$,
where $\mathbf 1 = P\mathbf 1$ is the \emph{right} eigenvector
corresponding to $\mathbf p$.

The left eigenvector of $P$ with $\lambda$ closest to 1, $\mathbf
l_2$, is a signed vector which decays at the slowest rate
\cite{Froyland-97, Pikovsky-Popovych-03}.  Sets where the magnitude
of the components of $\mathbf l_2$ maximize are the most dynamically
disconnected as a random walker starting there will take the longest
time to transit to other sets.  The corresponding right
eigenvector, $\mathbf r_2$, is also a signed vector, but it typically
includes well-defined plateaus.  A random trajectory conditioned
on starting in a set forming the support of a plateau is expected
to distribute in the long run, albeit only temporarily, as $\mathbf
l_2$ where it takes a single like sign \cite{Koltai-11, Froyland-etal-14}.
Such sets are therefore expected to form \emph{basins of attraction}
for time-asymptotic \emph{almost-invariant sets}.

Decomposition of the ocean flow into weakly disjoint basins of
attraction for almost-invariant attractors using the above spectral
method has been shown to form the basis of a \emph{Lagrangian
geography} of the ocean, where the boundaries between basins are
determined from the Lagrangian circulation itself, rather than from
arbitrary geographical divisions \cite{Froyland-etal-14, Miron-etal-17,
Miron-etal-18a}. The number of Lagrangian provinces will depend on
the number of right eigenvectors considered.  A large gap in the
eigenspectrum of $P$ provides a cutoff criterion for eigenvector
analysis, as provinces extracted from eigenvectors with eigenvalues
on the right of the gap will have significantly shorter retention
times than those extracted from eigenvectors on the left.

\section*{Results}

\paragraph{Deterministic analysis.}

The deterministic tools are applied on a velocity field derived
from sea-surface height (SSH) in the region of interest, spanning
$[70^{\circ}\text{W}, 50^{\circ}\text{W}] \times [55^{\circ}\text{S},
30^{\circ}\text{S}]$.  Specifically, we consider 
\begin{equation}
  v(x,t) = gf^{-1}\nabla^\perp\eta(x,t)
\end{equation}
where $g$ is gravity, $f$ is the mean Coriolis parameter, and
$\eta(x,t)$ is the superimposition on a mean dynamic topography
\cite{Rio-etal-11} of a daily $0.25^{\circ}$-resolution SSH anomaly
field constructed from along-track satellite altimetry measurements
\cite{LeTraon-etal-15}.  Such an altimetry-derived velocity reflects
an integral dynamic effect of the density field near the ocean
surface, i.e., above the thermocline \cite{Fu-etal-10}.

\begin{figure}[t!]
  \centering%
  \includegraphics[width=.5\textwidth]{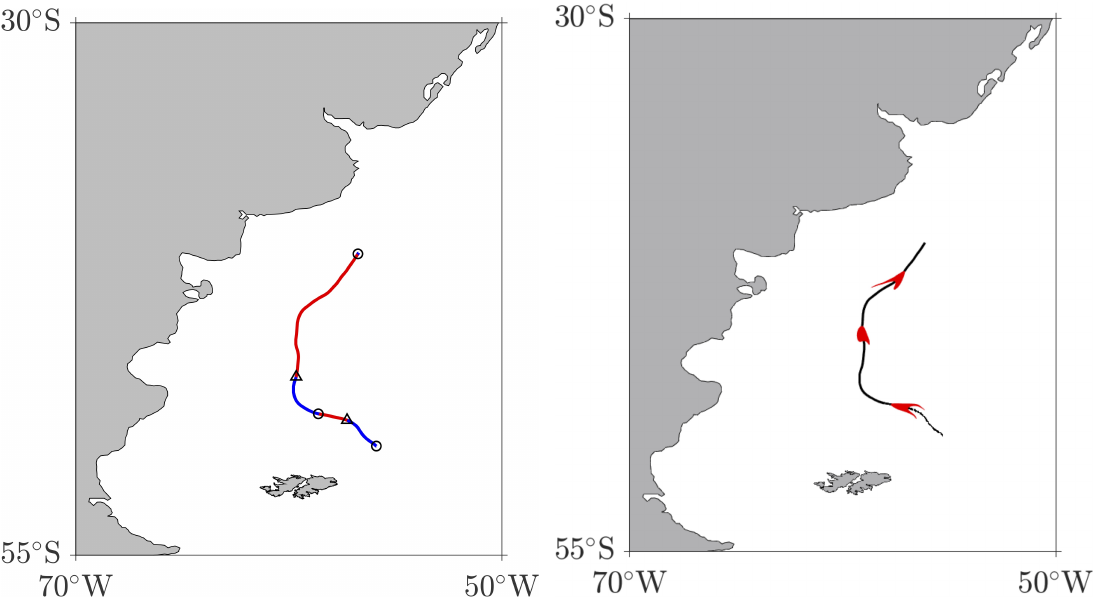}%
  \caption{(left) Shearless-parabolic LCS extracted from altimetry-derived
  velocity on $t_0 = \text{12 Dec 2001}$ using backward integration
  with $T = -15$ days.  Portions of the LCS colored red and blue
  are nearly neutral squeezing and stretching tensorline segments
  of the Cauchy--Green tensor field, respectively, connecting wedge
  and trisector singularities of the field, indicated by triangles
  and circles, respectively.  (right) Forward-advected images at
  $t_0$ of circles axially straddling the backward-advected image
  of the LCS at $t_0 - |T|$.}
  \label{fig:lcs}%
\end{figure}

Figure \ref{fig:lcs} illustrates the application of the
deterministic procedure on $t_0 = \text{12 Dec 2001}$ with $T = t
- t_0 = -15$ days, which was kept fixed throughout.  Here and in the
calculations that we describe below, the integration of trajectories
\eqref{eq:v} and tensorlines \eqref{eq:xi} are carried out using a
stepsize-adapting fourth-order Runge--Kutta method.  Unlike trajectory
integration, tensorline integration involves stepwise orienting the
eigenvector field(s).  Interpolations are obtained using a cubic
scheme.  Flow map differentiation in \eqref{eq:CG} is performed
using finite differences.  The Lagrangian coherence horizon $|T| =
15$ days was selected such that sufficiently long shearless-parabolic
LCS can be extracted.  Long LCS impact transport more effectively
than short LCS, which tend to coexist with a much longer LCS for
this $|T|$ choice and thus are ignored. As $|T|$ is increased,
singularities of the Cauchy--Green tensor tend to proliferate,
resulting typically only in very short shearless-parabolic LCS with
small effect on transport.  The (time) integration direction, in
turn, has been conveniently adopted to enable comparison with
observed (e.g., satellite-derived) tracer distributions.  More
specifically, a tracer distribution at any given time is the result
of the action of the flow on the tracer \emph{up to} that time.

Depicted in the left panel of Fig.\ \ref{fig:lcs} is the longest
shearless-parabolic LCS found in the domain.  The nearly neutral
squeezing and stretching Cauchy--Green tensorline segments that
form the LCS are shown in blue and red, respectively.  The wedge
and trisector singularities connected by these segments are indicated
by triangles and circles, respectively.  The shearless-parabolic
nature of the extracted LCS is demonstrated in the right panel of
Fig.\ \ref{fig:lcs}, which shows the forward-advected image at time
$t_0$ of three tracer disks axially straddling the backward-advected
image of the LCS at $t_0 - |T|$.  Note that the disks deform, as
expected, into V shapes which very closely axially straddle the LCS
at $t_0$.

\begin{figure}[t!]
  \centering%
  \includegraphics[width=.5\textwidth]{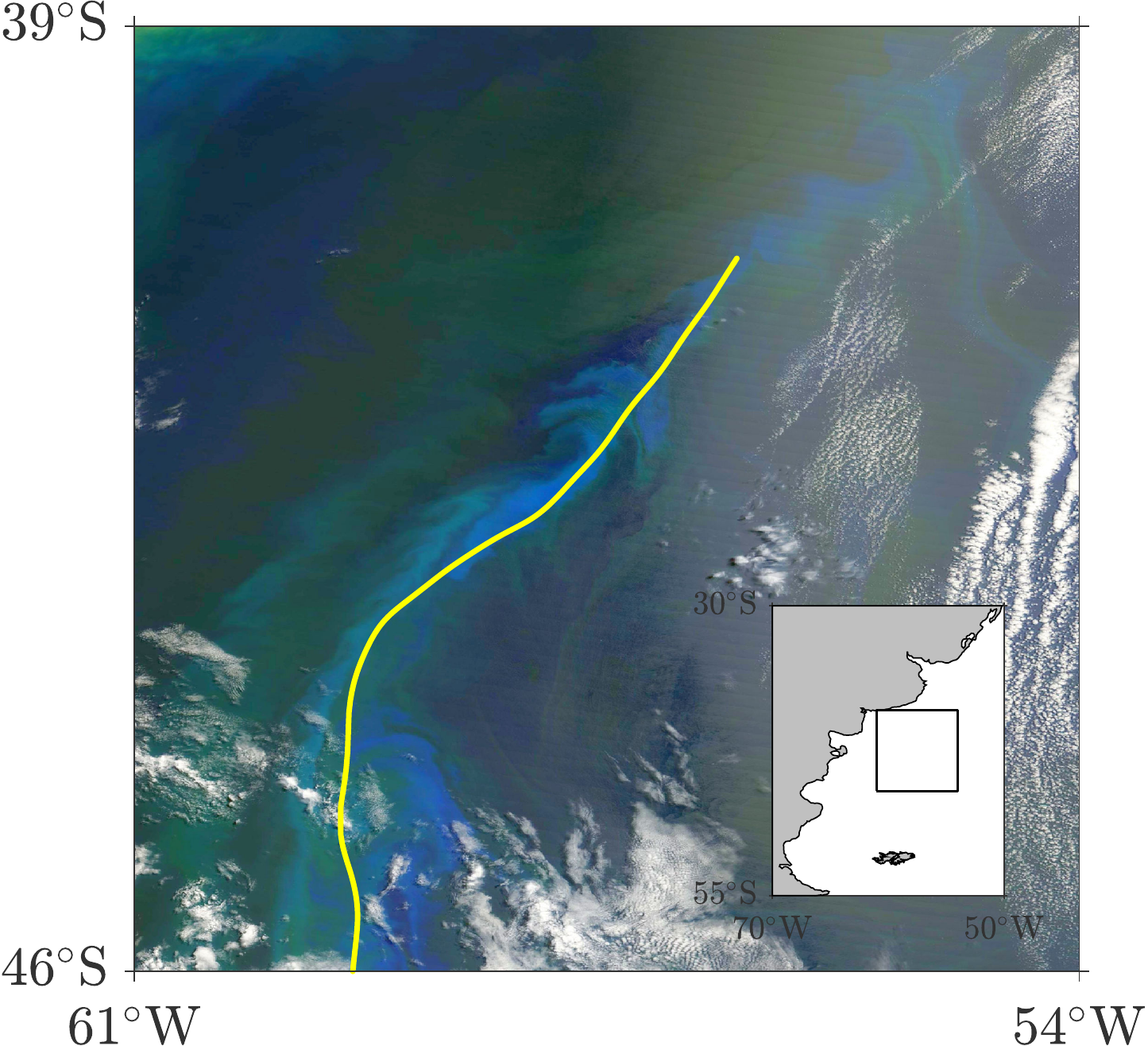}%
  \caption{Ocean pseudo color image on 12 Dec 2001 derived from the
  Moderate Resolution Imaging Spectroradiometer (MODIS) aboard
  \emph{Terra} with the shearless-parabolic LCS extracted from
  altimetry overlaid.}
  \label{fig:lcs-color}%
\end{figure}

Figure \ref{fig:lcs-color} shows a satellite-derived ocean pseudo-true
color image on $t_0 = \text{12 Dec 2001}$ with the extracted
shearless-parabolic LCS overlaid.  The coloration in the image is
determined by the interactions of incident light with particles
present in the water, mainly pigment chlorophyll, sediments, and
dissolved organic material.  Thus patterns formed in an ocean color
image can be thought, to first approximation, as developed by a
passive tracer.  Note the various V-shaped patches nearly axially
straddling the extracted LCS.  This provides strong independent
observational support for the altimetry-inferred LCS and its
shear-parabolic nature.  Ocean color images showing V shapes of the
type reported here are very rare; to the best of our knowledge we
report their occurrence for the first time.   Analogous V shapes,
dubbed ``chevrons,'' have been relatively recently observed in cloud
distributions in the weather layer of Jupiter at the boundaries
between so-called belts and zones organized around zonal jets
\cite{Simon-etal-12}. Jovian zonal jets have been rigorously
characterized as shear-parabolic LCS \cite{Hadjighasem-Haller-16}
and earlier heuristically as twistless KAM tori \cite{Beron-etal-08a}.

Shearless-parabolic LCS extraction was applied on sliding time
windows $[t_0-|T|,t_0]$ with $t_0$ selected every two weeks since
15 Oct 2002 until 15 Sep 2005.  During the period chosen, altimetry
measurements were made by altimeters mounted on four satellites,
thereby maximizing their availability and thus the quality of the
derived velocity \cite{Pascual-etal-06, Beron-etal-10b}.

\begin{figure}[t!]
  \centering%
  \includegraphics[width=.5\textwidth]{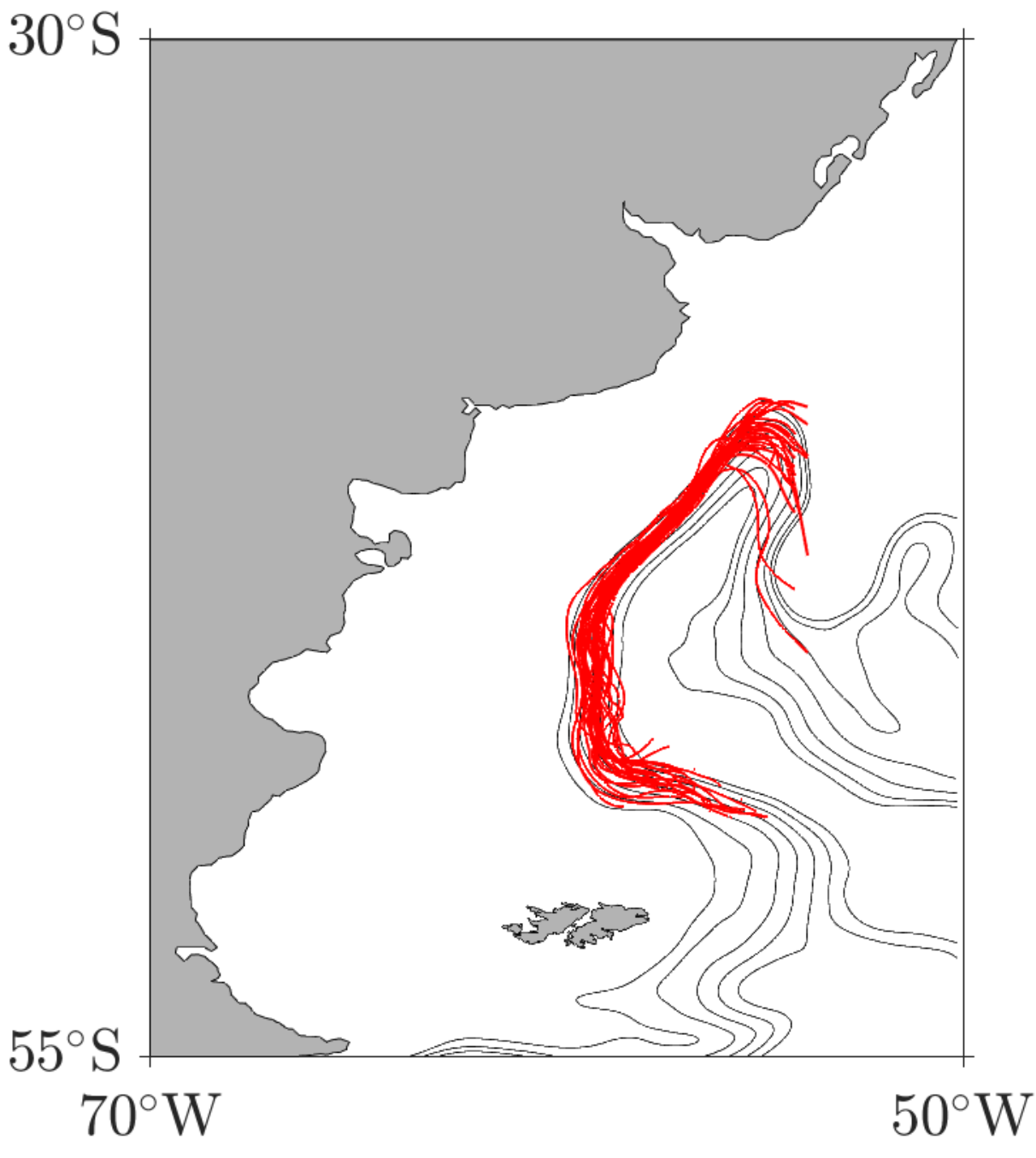}%
  \caption{Mean (15 Oct 2002--15 Sep 2005) streamlines along the
  Malvinas current core (black) overlaid with shearless-parabolic
  LCS extracted over windows $[t_0 + T,t_0]$ with $t_0$ sliding
  monthly over 30 Oct 2002--15 Sep 2005 and $T = -15$ days.}
  \label{fig:lcs-streamlines}%
\end{figure}

The extracted shearless-parabolic LCS are depicted (in red) in Fig.\
\ref{fig:lcs-streamlines} along with mean (over the LCS extraction
period, 15 Oct 2002--15 Sep 2005) streamlines of the altimetry-derived
flow.  The latter were selected to fill a strip around the Eulerian
axis of the Malvinas Current, here taken as the streamline where
the mean velocity maximizes at 42$^{\circ}$S.  Note that LCS and
streamlines do not coincide in position.  Yet they run close inside
the latitudinal band between about 38 and 49$^{\circ}$S.  This
suggests that the Malvinas Current behaves, within that latitudinal
band, as a quasi-steady shearless-parabolic LCS.  As such, it
inhibits cross-stream transport persistently over time, largely
preventing Patagonian shelf water from mixing with off-shelf water.
The rather tightly packed collection of LCS forms the Lagrangian
axis (or, more accurately, core) of the current.

\begin{figure}[t!]
  \centering%
  \includegraphics[width=.5\textwidth]{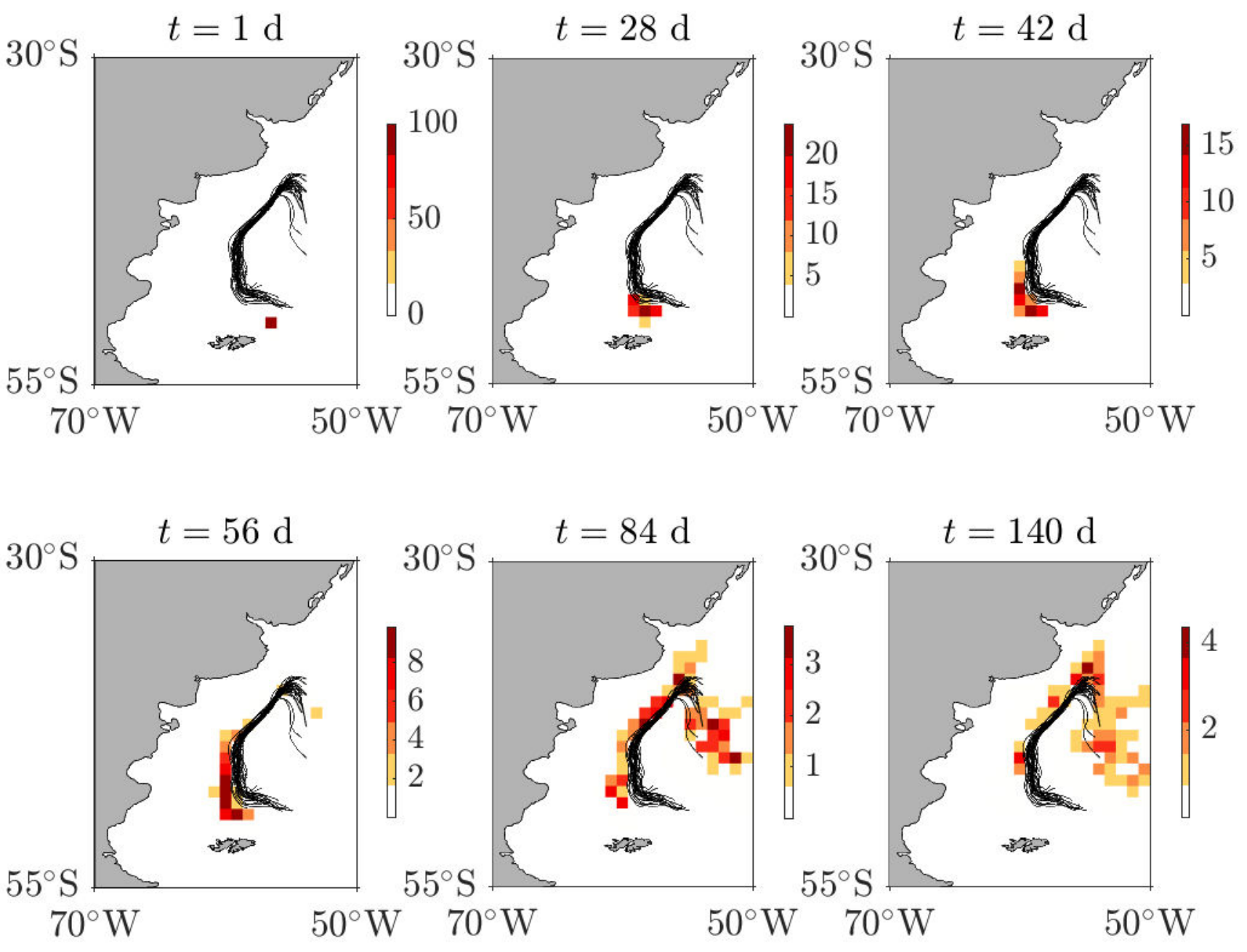}
  \caption{Snapshots of the ensemble-mean evolution over 15 Oct
  2002--15 Sep 2005 of tracers under the altimetry-derived flow
  with the extracted shearless-parabolic LCS overlaid.  The percentage
  of tracer particles visiting the boxes of a grid covering the
  domain is shown.}
  \label{fig:lcs-ensemble}%
\end{figure}

The cross-stream barrier nature of the Malvinas Current is verified
explicitly by the ensemble-mean evolution of tracers under the
altimetry-derived flow.  Selected snapshots are shown in Fig.\
\ref{fig:lcs-ensemble}.  The ensemble-mean tracer evolution was
computed by evolving the tracers from the same initial location on
the shelf northeast of the Malvinas Islands, every two months over
15 Oct 2002--15 Sep 2005, and then computing on a daily basis during
roughly half a year the percentage of tracer particles visiting
each 0.75$^{\circ}$-side box of a grid covering the domain.  Note
that the ensemble-mean tracer transport across the LCS is negligible.

Practically all of the transport off the shelf takes place near
38$^{\circ}$S, where the collection of extracted shearless-parabolic
LCS are interrupted, prior most of them turn very briefly eastward
and a few prolong a bit longer southeastward. The transport is
directed eastward and then mainly southeastward, out of the domain
through two exit routes, one at about 40$^{\circ}$S and another one
at 47$^{\circ}$S or so.  It is important to realize that this is
not obvious from the inspection of the mean streamlines, which
suggest mainly eastward transport for a tracer released on the shelf
at 38$^{\circ}$S, latitude at about which the Malvinas Current
encounters the Brazil Current \cite{Matano-93}.

The close proximity of the shearless-parabolic LCS and the mean
streamlines within 38 and 49$^{\circ}$S suggests KAM-like behavior.
In that latitudinal band, a decomposition of the flow into a steady
(reference) component and a small unsteady (perturbation) component,
certainly much smaller than near 38$^{\circ}$S where mesoscale
activity is known to be rather strong \cite{Goni-etal-96}, may be
envisioned as in earlier work \cite{Samelson-92}, in principle
enabling a near-integrable Hamiltonian system stability analysis
\cite{Arnold-etal-06}.  However, the flow is not recurrent, neither
in time nor in space.  In addition, quite unlike KAM curves, only
portions of the reference Hamiltonian level sets (mean streamlines)
are seen to ``survive'' under perturbation (i.e., when motion is
produced by the total flow).  These important differences indicate
that ocean jets can sustain robust barriers for transport far beyond
theoretical expectation \cite{Rypina-etal-07b}.

\paragraph{Probabilistic analysis.}

The probabilistic tools are applied on daily interpolated trajectories
produced by satellite-tracked drifting buoys from the NOAA Global
Drifter Program \cite{Lumpkin-Pazos-07} that have sampled the domain
of interest. The drifter positions are satellite-tracked by the
\emph{Argos} system or GPS (Global Positioning System). The drifters
follow the SVP (Surface Velocity Program) design, consisting of a
surface spherical float which is drogued at 15 m, to minimize wind
slippage and wave-induced drift \cite{Sybrandy-Niiler-91}.  The
drogue may not be present for the whole extent of a trajectory
record \cite{Lumpkin-etal-12, Beron-etal-16}.  We only consider
trajectory portions during which the drogue is present, so a
comparison with the altimetry-based results can be attempted.

We first cover the domain by 0.5$^{\circ}$-side boxes. The size
of the boxes was selected to maximize the grid's resolution while
each individual box is sampled by enough drifters. There are on
average 28 drifters per box independent of the day over 1993--2013.
The $(i,j)$th component of the transition matrix $P$ in \eqref{eq:Pij}
is estimated by counting the number of drifters which, visiting at
any time $t$ box $B_i$, enter box $B_j$ at $t + \mathcal T$, and
then dividing by the number of drifters in $B_i$.

\begin{figure}[t!]
  \centering%
  \includegraphics[width=.5\textwidth]{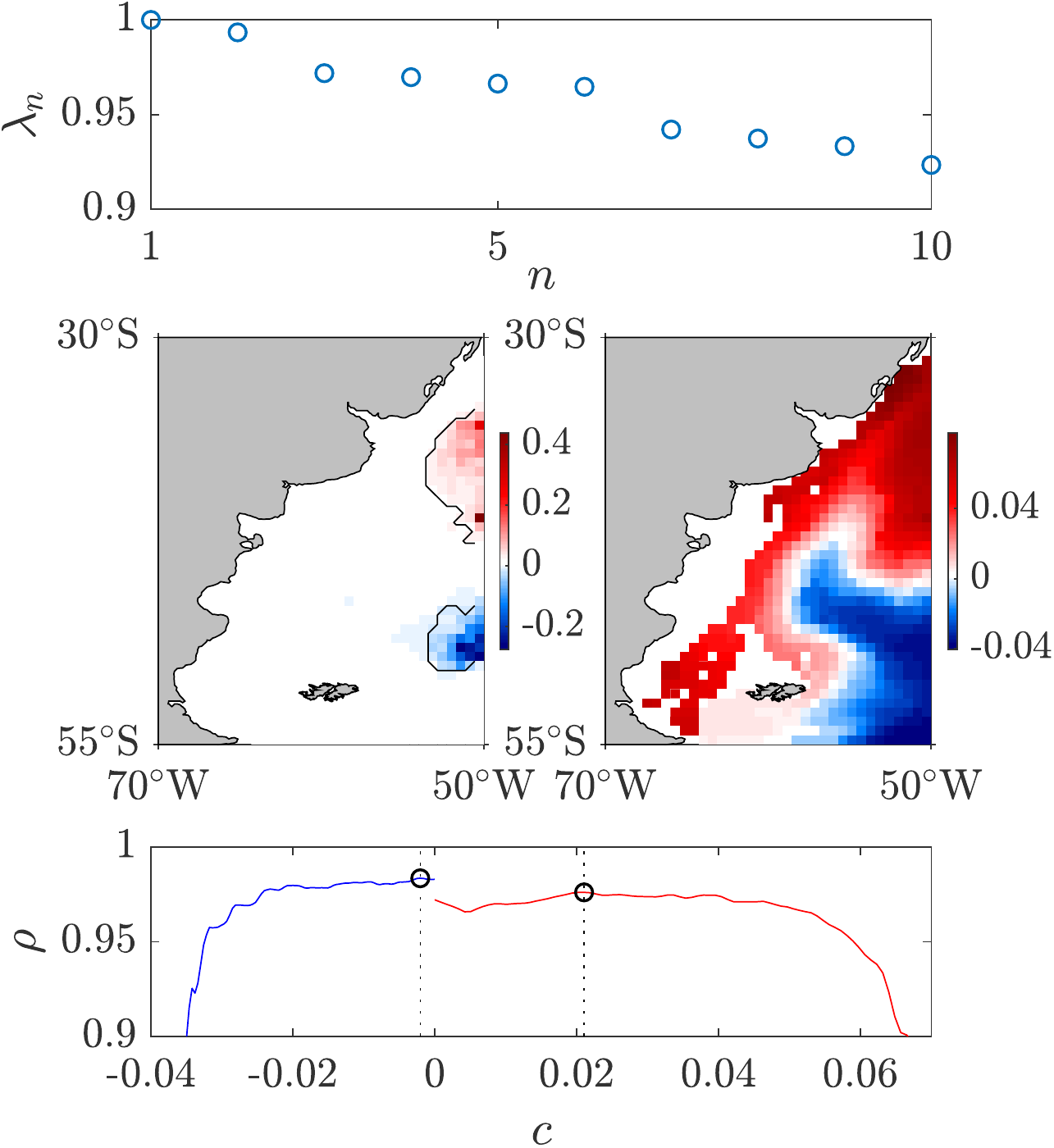}%
  \caption{(top) A portion of the discrete eigenspectrum of the
  drifter-based transition matrix $P$ showing the top 15 real
  eigenvalues. (middle) Left and right ($\mathbf r_2$) eigenvectors
  of $P$ with largest none unity eigenvalue ($\lambda_2 = 0.9946$).
  (bottom) Probability of a trajectory to be retained within regions
  where $\mathbf r_2 < c$ if $c < 0$ or $\mathbf r_2 > c$ if $c >
  0$ conditioned on starting in those regions.}
  \label{fig:geo-evec}%
\end{figure}

We have set $\mathcal T = 2$ days.  This in general guarantees
interbox communication.  Furthermore, $\mathcal T = 2$ days is
longer than the Lagrangian decorrelation timescale, which has been
estimated to be of about 1 day \cite{LaCasce-08}.  Markovian
dynamics can be expected to approximately hold as there is negligible
memory farther than 2 days into the past.  A similar reasoning was
applied in earlier applications involving drifter data
\cite{Maximenko-etal-12, vanSebille-etal-12, Miron-etal-17,
McAdam-vanSebille-18, Miron-etal-18a, Olascoaga-etal-18}. Here
the validity of the Markov model was estimated by checking that
$\lambda(P(n\mathcal T)) = \lambda(P(\mathcal T))^n$ holds well
with $n$ up to 5 and consistent with this we have verified that the
results presented below are largely insensitive to variations of
$\mathcal T$ in the range 2--10 days.

We note that while the domain is open, $P$ has been constructed in
such a way that it is row-stochastic by excluding all drifter
trajectory pieces, which, starting inside the domain, terminate
outside.  It must be emphasized that this does not force trajectories
to spuriously bounce back into the domain.  The signature of inward
motion is imprinted in the drifter trajectory data, so is in the
resulting Markov-chain model.  On the other hand, working with a
row-stochastic $P$ facilitates the interpretation of the probabilistic
tool results, albeit clearly not without exerting some care.  Applying
the Tarjan algorithm \cite{Tarjan-72} on the directed graph
associated with the corresponding Markov chain reveals the existence
of a set of boxes in the southwestern corner of the domain that are
not reachable from boxes in its complement.  The constructed $P$
is thus reducible.  Nevertheless, the complement of that set of
boxes covers most of the domain and furthermore is absorbing.  So
excluding it to make $P$ irreducible is inconsequential.

With the above in mind, we show in the top panel of Fig.\
\ref{fig:geo-evec} a portion of the eigenspectrum of $P$ corresponding
to the largest 10 real eigenvalues.  The largest eigenvalue equals
unity and is simple.  Consequently, the associated left eigenvector,
which we loosely refer to as $\mathbf p$, is invariant, yet it is
not strictly positive. The right eigenvector is $\mathbf 1$. Any
probability vector forward evolves under left multiplication by $P$
into $\mathbf p$, whose components maximize along the eastern
boundary of the domain.  More specifically, this happens inside the
regions delimited by the black curves in the middle-left panel of
Fig.\ \ref{fig:geo-evec}. A tracer, irrespective of how it is
initialized in the domain, will thus in the long run accumulate in
those regions of the eastern boundary.  Physically this means that
it will eventually exit the domain through those locations. Once
the tracer gets attracted there (exists the domain) it will not
recirculate back into the domain.

\begin{figure}[t!]
  \centering%
  \includegraphics[width=.5\textwidth]{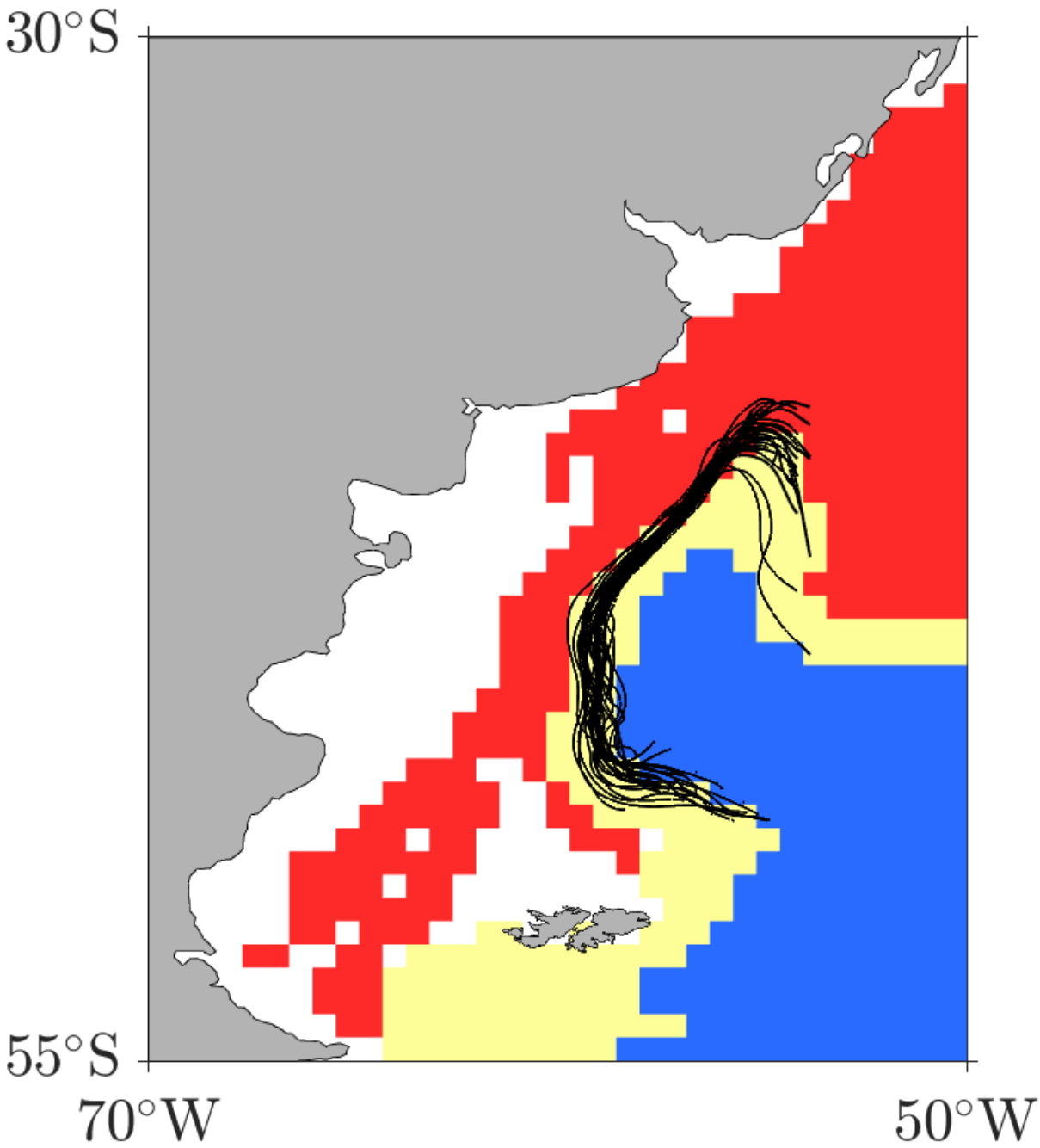}%
  \caption{Lagrangian geography deduced from the structure of the
  dominant right eigenvector of the drifter-based transition matrix
  $P$ overlaid with all shearless-parabolic LCS extracted from
  satellite altimetry.  Retention probability and time are maximized
  in the red and blue provinces, which represent the main Lagrangian
  provinces, while the yellow province represents a transition
  region with smaller retention probability and time (cf.\ Table
  \ref{tab:geo}).}
  \label{fig:geo}%
\end{figure}

The middle panel of Fig.\ \ref{fig:geo-evec} shows the left ($\mathbf
l_2$) and right ($\mathbf r_2$) eigenvectors of $P$ corresponding
to the second eigenvalue closest to 1 ($\lambda_2 = 0.9934$).
Note the two regions where the magnitude of the components $\mathbf
l_2$ maximize.  The support of these small regions represent
almost-invariant attracting sets for tracers initially distributed
on the large regions where $\mathbf r_2$ takes constant values,
which represent their basins of attraction.  (The eigenvectors have
been arbitrary assigned sings such that regions of $\mathbf r_2$
evolve to like signed regions of $\mathbf l_2$.) These almost-invariant
attracting sets, centered at about 38.5 and 48$^{\circ}$S at the
eastern side of the domain, physically represent routes of escape
out of the domain for tracers in the corresponding basins of
attraction.  Traces initially inside each basin will passively
evolve toward the respective attractor, which, being almost invariant,
will retain the tracers temporarily until they are eventually drained
out of the domain.  Thus while $\mathbf p$ indicates that tracers
will eventually exit the domain through the eastern boundary,
$\mathbf l_2$ reveals preferred exit paths depending on how they
are initialized.

The eigenspectrum of $P$ in the portion shown in the top panel of
Fig.\ \ref{fig:geo-evec} reveals a gap between $\lambda_2$ and
$\lambda_3$.  Indeed, there is a drop of 2.1702\pct{ }from $\lambda_2$
to $\lambda_3$, while $\lambda_1$ and $\lambda_2$ only differ by
0.0066\pct{ }and $\lambda_3$ through $\lambda_5$ are very similar,
changing by just 0.0091\pct{ }on average. This suggests that a
minimal significant Lagrangian geography with sufficiently large
weakly communicating provinces to substantively constrain connectivity
in the domain can be constructed by inspecting $\mathbf r_2$.  A
geography composed of smaller and less isolated provinces may be
obtained by inspecting additional right eigenvectors, but this is
not pursued here as our interest is to independently verify the
deterministic analysis of the altimetry data, which suggested weak
communication between shelf water on the west of the Malvinas Current
and the open-ocean water on the east.

\begin{table}
  \centering
  \begin{tabular}{lccc}
  \hline
         & red     & yellow & blue\\
  $\rho$ & 0.9763  & 0.7797 & \,\,\,0.9835\\ 
  $\tau$ & 7.7952 & 1.1518 & 11.4850\\
  \hline
  \end{tabular}
  \caption{Retention probability ($\rho$) and time ($\tau$, in
  months) inside each of the three provinces of the Lagrangian
  geography in Fig.\ \ref{fig:geo}, which are labelled by their
  colors in the figure.}
  \label{tab:geo}
\end{table}

Shown in Fig.\ \ref{fig:geo} is the constructed minimal Lagrangian
geography.  It includes three provinces, which are defined as
follows.  Rather than defining the Lagrangian provinces as sets
where $\mathbf r_2$ takes one sign, as done in earlier work
\cite{Froyland-etal-14, Miron-etal-17, Miron-etal-18a}, here we
define them as sets where the retention probability is maximized.
More specifically, let $\mathcal A \subset \{1, \dotsc, N\}$ and
define $A := \smash{\bigcup_{i\in \mathcal A}} B_i \subset X$. If
one conditions on a tracer trajectory to start in set $A$, the
probability to be retained within $A$ after one application of $P$
is $\rho(A) := \smash{\sum_{i,j\in \mathcal A}}p_iP_{ij}/\smash{\sum_{i\in
\mathcal A}} p_i$ \cite{Dellnitz-Junge-99, Froyland-05}.  The bottom
panel shows $\rho$ for $A(c) := \smash{\bigcup_{i: \mathbf r_2 <
c}} B_i$ if $c < 0$ or $\smash{\bigcup_{i: \mathbf r_2 > c}} B_i$
if $c > 0$.  We compute $\max_c\rho(A(c)) = 0.9835$ and $0.9763$
at at $c = c_- = -0.0016$ and $c = c_+ = 0.0219$, respectively. The
sets $A_\text{blue} := \smash{\bigcup_{i: \mathbf r_2 < c_-}} B_i$
and $A_\text{red} := \smash{\bigcup_{i: \mathbf r_2 > c_+}} B_i$,
depicted red and blue in Fig.\ \ref{fig:geo}, respectively, form
the main Lagrangian provinces.  The set depicted yellow, $A_\text{yellow}
:= \smash{\bigcup_{i: c_- < \mathbf r_2 < c_+}} B_i$, represents a
transition province with smaller retention probability,
$\rho(A_\text{yellow}) = 0.7797$.

Larger (smaller) retention probability is associated with longer
(shorter) retention time.  A simple measure of retention time is
computed as follows.  Consider $\mathbf p_A P|_A = \lambda_A \mathbf
p_A$, where $P|_A$ is $P$ restricted to some set $A\subset X$ and
$\lambda_A$ is the largest eigenvalue of $P|_A$.  If $P$ is
irreducible, $\lambda_A < 1$ and $\mathbf p_A \ge 0$.  Assume that
a tracer trajectory starts in $A$.  If the trajectory is conditioned
on being retained in $A$, it will asymptotically distribute as
$\mathbf p_A$, where $\mathbf p_A$ has been normalized to a probability
vector.  Such a $\mathbf p_A$ is called a quasi-stationary distribution
(cf.\ Chapter 6.1.2 of Bremaud \cite{Bremaud-99}).  The expectation
of the random time to exit $A$ is $\tau(A) := \mathcal T/(1-\lambda_A)$
(cf.\ Section B.7 in the online Supplementary Information).  Such
a $\tau(A)$ provides an average measure of retention time in $A$.
We compute $\tau(A_\text{blue}) = 11.5$ months and $\tau(A_\text{red})
= 7.7952$ months for the main Lagrangian provinces, and a much
shorter retention time, $\tau(A_\text{yellow}) = 1.1518$ months,
for the transition province.

\begin{figure}[t!]
  \centering%
  \includegraphics[width=.5\textwidth]{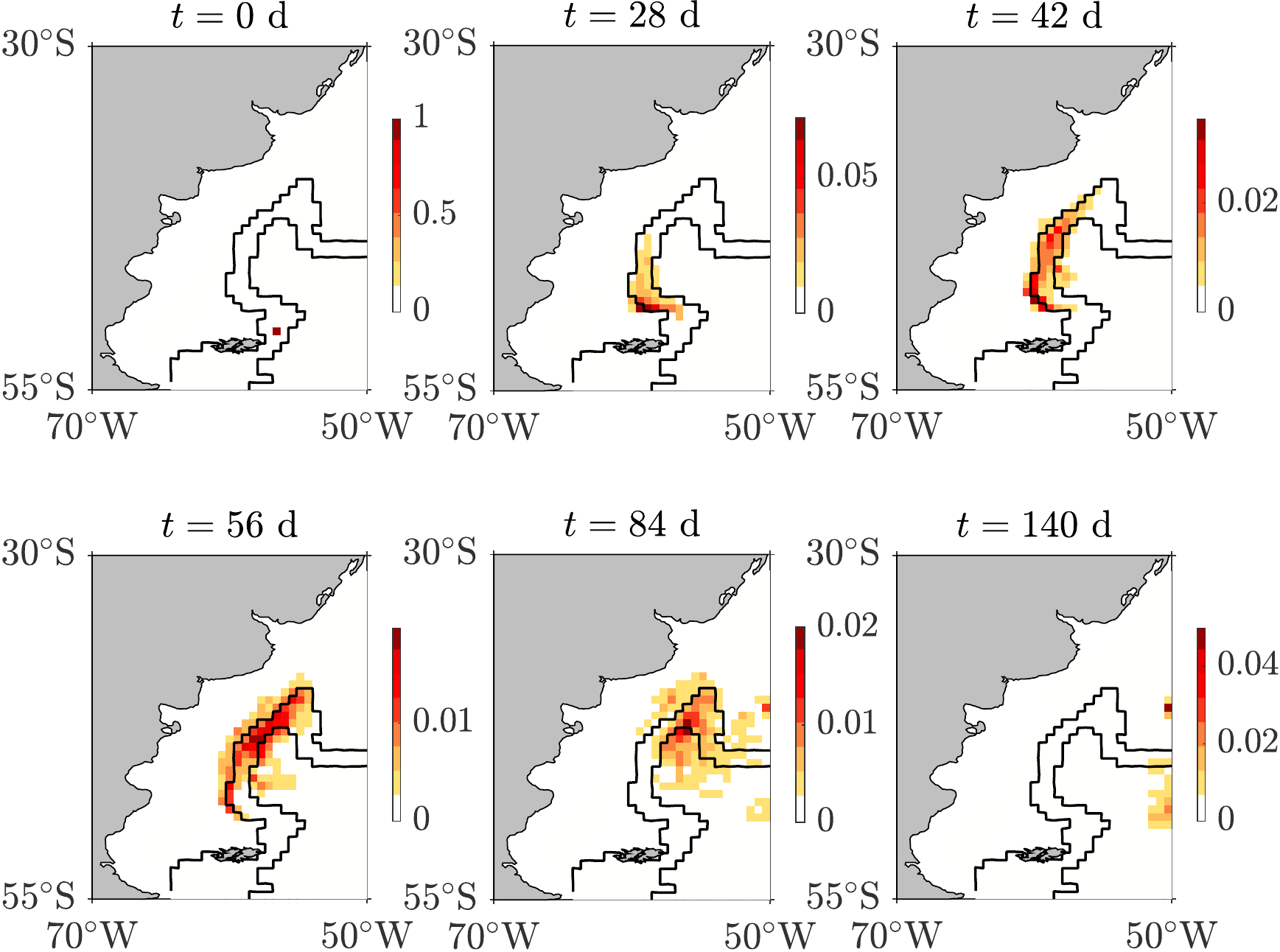}%
  \caption{Snapshots of the drifter-based forward evolution of a
  probability density with the boundaries of the main Lagrangian
  provinces indicated.}
  \label{fig:geo-fwd}%
\end{figure}

Clearly, the partition of the flow domain provided by the drifter-based
Lagrangian geography is indicative of low connectivity between shelf
water and open-ocean water off the shelf, south of 38$^{\circ}$S.
Figure \ref{fig:geo-fwd} provides confirmation for this inference
from direct calculation.  More specifically, this figure shows
selected snapshots of the evolution under left multiplication by
$P$ of a tracer probability initially on the shelf, northeast of
the Malvinas archipelago.  Note that up to day 56, the tracer
probability propagates northeastward, predominantly confined within
the transition province of the Lagrangian geography.  The
almost-invariant character of the boundaries of the Lagrangian
provinces explains the small leakage of probability over the main
provinces east and west of the transition province.

The leakage continues past day 56 of evolution, becoming stronger
as the Brazil--Malvinas Confluence near 38$^{\circ}$s is reached.
Time-asymptotically the probability that leaks to the west and east
of the transition province accumulates in the southern and northern
almost-invariant attractor, respectively (cf.\ Fig.\ \ref{fig:geo-evec},
middle-left panel).  These attractors, we reiterate, physically
represent exit routes out of the domain.

We note that if the tracer probability were initialized inside the
transition province of the Lagrangian geography where this turns
(south)eastward at about 38$^{\circ}$s, it would remain confined
within for a shorter period of time before leaking out and being
absorbed into the nearest almost-attracting set.  The reason for
this is the much closer proximity of the attractors to the transition
province at these latitudes.  The retention time measure
$\tau(A_\text{yellow}) = 1.1518$ months, discussed above, is an average
measure for the entire transition province.  The average retention
time within the portion of transition province that lies (roughly)
inside the Brazil--Malvinas Confluence region is somewhat smaller
than 1 week.  By contrast, the average retention time in the
complement of this set is nearly 5 weeks, which is very close to
the average retention time in the whole transition province.  This
is consistent with the behavior just described. Also consistent
with this is the strong mesoscale variability that affects the area
where the Malvinas and Brazil currents meet.  Diffusion is benefited
from such variability, which contributes to shorten the retention
time there.

\begin{figure}[t!]
  \centering%
  \includegraphics[width=.5\textwidth]{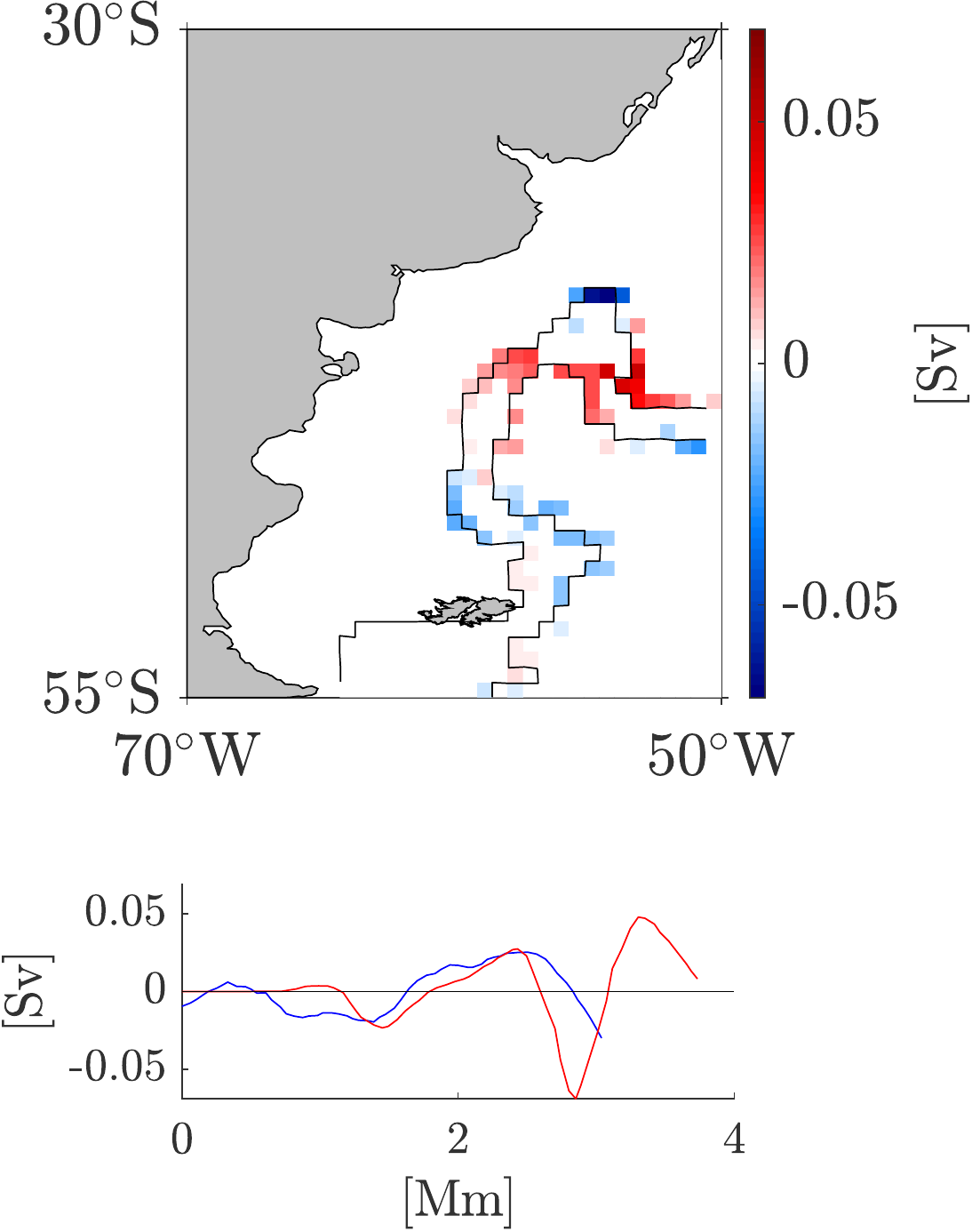}%
  \caption{(top) Drifter-based estimate of the flux across the
  boundaries of the transition province of the Lagrangian geography.
  (bottom) The flux shown as a function of arclength (increasing
  northward) along each boundary.  The red (blue) curve corresponds
  to the boundary with the main province painted red (blue) in Fig.\
  \ref{fig:geo}.}
  \label{fig:geo-flx}%
\end{figure}

To assess the latter, one can leverage on the computation of the
flux across the boundary of a set $A (= \smash{\bigcup_{i\in\mathcal
A}} B_i)$, which is readily accomplished as follows
\cite{Dellnitz-etal-09}.  Let $\partial\mathcal A (\subset \mathcal
A)$ be the index set of boxes on the boundary of $A$.  The flux
$\Phi(B_j)$ through a boundary box $B_j$, with $j \in \partial\mathcal
A$, can be calculated as $\Phi(B_j) = \Phi_\mathrm{out}(B_j) -
\Phi_\mathrm{in}(B_j)$, where $\Phi_\mathrm{out}(B_j) = \smash{\mathcal
T^{-1}} \cdot \vol(B_j) \smash{\sum_{k\in\{1,\dotsc,N\}\setminus
\mathcal A}} P_{jk}$ and $\Phi_\mathrm{in}(B_j) = \smash{\mathcal
T^{-1}} \cdot \smash{\sum_{k\in\{1,\dotsc,N\}\setminus \mathcal A}}
\vol(B_k) P_{kj}$.  Figure \ref{fig:geo-flx} shows an evaluation
of the flux formulas for the transition province ($A_\text{yellow}$),
with $\vol(B_i)$ estimated as $\area(B_i) \cdot H$ where $H = 15$
m is the drogue depth.  Note that the flux through the boundary
boxes of these sets tend to maximize inward or outward in the
Brazil--Malvinas Confluence region.

We note finally that despite the limitation provided by the number
of drifters available, particularly over the continental shelf, the
inferred Lagrangian provinces are in very good agreement with the
biophysical provinces deduced by Longhurst \cite{Longhurst-98} and
more recently by Saraceno et al. \cite{Saraceno-etal-05, Saraceno-etal-06}
using two independent methods.

\paragraph{Synthesis of deterministic and probabilistic
analyses.}

The results of the deterministic analysis of the altimetry data and
the probabilistic analysis of the drifter data are largely consistent.
They both independently indicate Lagrangian stability for the
Malvinas Current, which largely behaves as a barrier that prevents
shelf water to its west from being mixed with off-shelf water to
its east.  This is well demonstrated in Fig.\ \ref{fig:geo}, which
shows that the shearless-parabolic LCS extracted from altimetry
over sliding time windows along the multiyear record analyzed lie
well within the transition province between the main Lagrangian
provinces constructed from all available drifter data.

The off-shelf transport takes place near 38$^{\circ}$S, where the
Malvinas Current meets the Brazil Current (cf.\ Figs.\
\ref{fig:lcs-ensemble} and \ref{fig:geo-fwd}).  This region is
characterized by strong mesoscale variability, which the probabilistic
analysis showed to promote diffusion in the region.  Consistent
with this, the deterministic analysis revealed LCS prolonging only
briefly southeastward at the Brazil--Malvinas Confluence latitude,
thereby allowing unrestrained exchanges there.  According to both
the deterministic and probabilistic results, the off-shelf export
eventually reaches the South Atlantic's interior through two
routes.

\paragraph{Lagrangian--Eulerian stability duality.}

The reported Lagrangian stability of the Malvinas Current motivates
the question of its stability in the Eulerian frame.  A stability
result for a general meandering meridional current with vertical
shear is lacking.  Yet the stability of a basic flow (steady solution)
$V(x,z)$ in thermal-wind balance, where $x$ is cross-stream and
$-z$ depth, of the $y$-independent, inviscid, unforced, nonhydrostatic,
Boussinesq equations on an $f$ plane is well established
\cite{Cho-etal-93, Mu-etal-96}.  Both sufficient and necessary
conditions for the symmetric stability of $V(x,z)$ under arbitrarily
large and shaped perturbations are given by $N^2/(\partial_z V)^2
> 1/(1 + \partial_x V/f) > 0$, where $N^2$ is the square of the
basic flow's Brunt--V\"ais\"al\"a frequency.  Note that symmetric
stability requires both static stability ($N^2 > 0$) as well as
inertial stability ($f^2 + f\partial_x V) > 0$).  Assuming stable
stratification, these conditions are equivalent to $fQ > 0$, where
$Q := N^2(f + \partial_x V) - f(\partial_z V)^2$ is the basic flow's
potential vorticity, which is materially preserved.  Clearly, $fQ
< 0$ is both necessary and sufficient for symmetric instability.
This condition includes the necessary condition for instability
under infinitesimally small normal-mode perturbations originally
derived by Hoskins \cite{Hoskins-74} (cf.\ Section C in the online
Supplementary Information for a review of the results just described).

\begin{figure}[t!]
  \centering%
  \includegraphics[width=.5\textwidth]{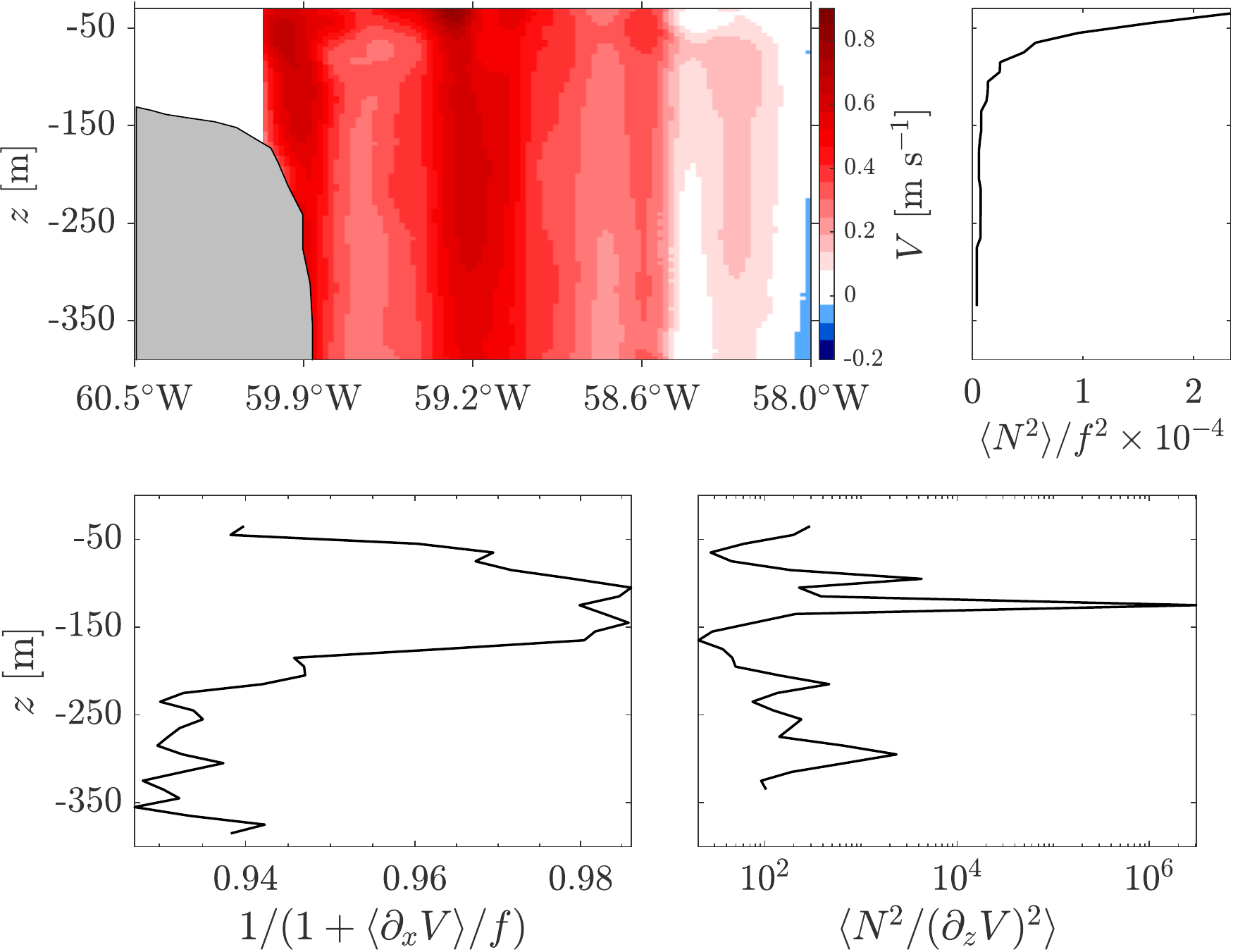}%
  \caption{(top left) Meridional velocity section along 45$^{\circ}$S
  across the Malvinas Current from a hull mounted ADCP collected
  on 28 Dec 1992 from RSS \emph{Discovery} during WOCE cruise A11.
  (top right) Normalized by the Coriolis parameter squared,
  along-transect average of Brunt-V\"ai\"sala frequency squared
  computed using temperature and salinity from CTD casts during
  WOCE A11. (bottom left) Along-transect-average of the ratio of
  the Coriolis parameter and the vertical component of the absolute
  vorticity.  (bottom right) Along-transect-average of Richardson
  number.}
  \label{fig:sym}%
\end{figure}

Using available direct high-resolution velocity measurements and
temperature and salinity data collected by RSS \emph{Discovery} in
late December 1992 during WOCE cruise A11 along 45$^{\circ}$S
\cite{Saunders-King-95}, we proceed to check if the Malvinas Current
has any hope to be symmetrically stable.  The WOCE-A11 transect
lies across the Malvinas Current, which we assume to be represented
as an along-stream-symmetric baroclinic parallel flow.  The velocity
data was collected by a hull mounted acoustic Doppler current
profiler (ADCP) in a westward course.  A section of meridional
(nearly along-stream) velocity is shown in the top-left panel of
Fig.\ \ref{fig:sym}.  The temperature and salinity were obtained
from conductivity-temperature-depth (CTD) casts occupied in a
returned eastward course; the Brunt--V\"ais\"al\"a frequency,
averaged along the section, is shown in the top-right panel Fig.\
\ref{fig:sym}.  The bottom-left and bottom-right panels of Fig.\
\ref{fig:sym} show along-section-mean $1/(1 + \partial_x V/f)$ and
$N^2/(\partial_z V)^2$, respectively.  Note that the symmetric
stability conditions are well satisfied on average across the
Malvinas Current. This result together with those from the deterministic
and probabilistic nonlinear dynamics analysis suggest a
Lagrangian--Eulerian stability duality for the current.

The above result is not obvious whatsoever.  Indeed, it is well-known
that unsteady laminar Eulerian flows can support irregular Lagrangian
motion (chaotic Lagrangian motion generically in bounded, recurrent
unsteady two-dimensional flows) \cite{Aref-84}. But there are several
caveats to have in mind.  First, a priori conditions for
stability/instability should be verified by the basic flow rather
than the total flow and instantaneously as we have checked here.
Yet Piola et al. \cite{Piola-etal-13} note that the ADCP velocity
shear in the 100- through 390-m depth interval is virtually identical
to the geostrophic shear derived from hydrography.  This suggests
that the ADCP velocity may be providing a reasonable representation
of the basic velocity.  Also, hydrographic data are not available
with enough longitudinal resolution to check the symmetric stability
conditions pointwise.  And last but not least, there are not
sufficient data to assess the extent to which along-stream symmetry
holds for the Malvinas Current. This has most chances to be verified
within 38--49$^{\circ}$S, (co)incidentally where shearless-parabolic
LCS and the mean streamlines were found to lie closest together.

\section*{Summary and final remarks}

In this paper we have characterized the Malvinas Current as an
enduring cross-stream transport barrier by applying nonlinear
dynamics tools of two quite different types on independent datasets.

One type of tools used was deterministic, built on a geometric,
objective (i.e., observer-independent) notion of material shear.
These tools were applied on velocities derived from satellite
altimetry, and revealed---for the first time from this dataset---Lagrangian
coherent structures (LCS) of the shearless-parabolic class.  Computed
over sliding time windows along a multiyear period of satellite
altimetry data with the highest density, the shearless-parabolic
LCS were found to form an enduring near-surface Lagrangian axis for
the Malvinas Current that largely inhibits shelf water on its western
side from mixing with open-ocean water on its eastern side.

The other type of nonlinear dynamics tools employed was probabilistic,
built on ergodic theory and describing tracer motion on a Markov
chain.  These were applied on available satellite-tracked drifter
trajectories, revealing statistically weak communicating Lagrangian
provinces separated by the LCS extracted from altimetry. This
provided independent support for the enduring role of the Malvinas
Current in the near surface as a cross-stream transport barrier.

The shear-parabolic nature of the Lagrangian axis of the Malvinas
Current was supported on satellite-derived ocean color imagery.
This revealed---for the first time to the best of our knowledge---V
shapes nearly axially straddling current's Lagrangian axis.  Similar
V shapes, referred to as ``chevrons,'' have been been relatively
recently observed in clouds distributions in the weather layer of
Jupiter, confirming the enduring nature of zonal jets there as
barriers for meridional transport.

In-situ velocity and hydrographic data showed that conditions for
symmetric stability are satisfied. This suggested a Lagrangian--Eulerian
stability duality for the Malvinas Current, a nonobvious result given
the known ability of laminar Eulerian flows to support irregular
Lagrangian motion.

The gas giant's chevrons have been connected to inertia--gravity
wave motion \cite{Simon-etal-12}.  Satellite imagery has recently
revealed internal waves propagating along the Patagonian shelfbreak
and continental slope in the opposite direction of the Malvinas
Current \cite{Magalhaes-daSilva-17}.  The possible connection with
the chevrons observed along the Lagrangian axis of the current
deserves to be investigated.  This is beyond the scope of this paper
as also is investigating how representative the results here presented
are of other western boundary currents.  For instance, high-resolution
measurements across the Gulf Stream suggest that symmetric stability
is violated locally along submesoscale fronts in the upper ocean
\cite{Thomas-etal-16}, which already indicates a potentially important
difference with the Malvinas Current.

\section*{Acknowledgments}

We thank Peter Koltai for the benefit of discussions on transfer
operators defined using stochastic kernels and Markov chains, and
Daniel Karrasch for the benefit of discussions on line fields.  We
also thank Joaquin Tri\~nanes for producing the MODIS-\emph{Terra}
ocean psuedocolor image in Fig.\ \ref{fig:lcs-color}.  MODIS-\emph{Terra}
data are available from NASA's OceanColor Web (https://\allowbreak
oceancolor.\allowbreak gsfc.nasa.\allowbreak gov) with support from
the Ocean Biology Processing Group (OBPG) at NASA's Goddard Space
Flight Center. The altimeter products were produced by SSALTO/DUCAS
and distributed by AVISO with support from CNES (http://\allowbreak
www.aviso.\allowbreak oceanobs). The drifter data are available
from the NOAA Global Drifter Program (http://\allowbreak
www.aoml.noaa.gov/\allowbreak phod/dac). This work has been supported
by ONR Global grant 12275382.

\section*{Author Contributions}

FJBV performed the probabilistic dynamical system analysis and the
Eulerian stability analysis, and wrote the manuscript; NB carried
out the deterministic dynamical system analysis; MS compiled the
hydrographic data used in the Eulerian stability analysis; MJO
compiled the drifter trajectory data involved in the probabilistic
dynamical system analysis; and FJBV, MS, and CS supervised NB's
work.  All authors contributed to the interpretation of the results
and reviewed the manuscript.

\section*{Additional Information}

\noindent\textbf{Competing interests:} The authors declare
no competing interests.

\section*{Online Supplementary Information for ``Lagrangian stability of
the Malvinas Current''}

\noindent The Supplemtary Information includes three appendices providing
additional details on the deterministic (A) and probabilistic (B)
tools employed in the paper as well as on the Eulerian stability
result considered (C).  This is done with a goal in mid of making
the paper sufficiently selfcontained.

\appendix

\numberwithin{equation}{section}

\section{Deterministic tools}

\subsection{Flow map and Cauchy--Green strain tensor}

Consider the motion equation for fluid particles,
\begin{equation}
  \dot{x} = v(x,t),
  \label{eq:dxdt}
\end{equation}
where $v(x,t)$ is a two-dimensional incompressible velocity field.
Solving this equation for fluid particles at positions $x_0$ at
time $t_0$, one obtains a map that takes the particles to positions
$x$ at a later time $t$, namely, $F_{t_0}^t(x_0) := x(t;x_0,t_0)$.

A fundamental objective measure of fluid deformation is given by
the Cauchy--Green (CG) strain tensor,
\begin{equation}
  C_{t_0}^t(x_0) := \D{F}_{t_0}^t(x_0)^\top
  \D{F}_{t_0}^t(x_0).
  \label{eq:C}
\end{equation}
Its eigenvalues and eigenvectors satisfy $0 < \lambda_1(x_0)\equiv
\lambda_2(x_0)^{-1} \leq 1$ and $\xi_1(x_0) \perp \xi_2(x_0)$,
respectively.  

\subsection{Squeezelines, stretchlines, and singularities}

It immediately follows that a well-behaved material curve that at
$t_0$ is everywhere tangent to $\xi_1$ (resp., $\xi_2$) will pointwise
squeeze (resp., stretch) over $[t_0,t]$ by $|\D{F}_{t_0}^t(x_0)\xi_1(x_0)|
= \smash{\sqrt{\lambda_1(x_0)}} < 1$ (resp.,
$|\D{F}_{t_0}^t(x_0)\xi_2(x_0)|$ $= \smash{\sqrt{\lambda_2(x_0)}}
> 1$). 

Good behavior of such \emph{squeezelines} and \emph{stretchlines},
i.e., solution curves $s\mapsto r(s)$ to
\begin{equation}
  r' = \xi_i(r), \quad i=1,2,
  \label{eq:drds}
\end{equation}
is guaranteed away from singular points $x_*$ where the CG tensor
is isotropic.  Singular points $x_*$ are such that $\lambda_1(x_*)
= \lambda_2(x_*)$ (or, equivalently in the incompressible case,
$C_{t_0}^t(x_*) = \Id$) and thus $\xi_1(x_*)$ and $\xi_2(x_*)$ take
arbitrary orientations. 

Singularities of planar eigenvector fields are analogous to stationary
points of planar vector fields \cite{Delmarcelle-Hesselink-94,
Kratz-etal-13}.  Unlike vector fields, eigenvector fields are
bidirectional.  However, curves tangent to them or \emph{tensorlines}
are well defined (away from singularities) and thus are referred
to as line fields.  Line fields on the plane admit two types of
structurally stable\footnote{Indeed, there can be singularities
with indices \cite{Spivak-99} other than $-\frac{1}{2}$ (trisector)
or $+\frac{1}{2}$ (wedge). But none of these are stable under
perturbations to the line fields: they will usually fall apart into
wedges and trisectors.  These, in turn, cannot be perturbed away.}
singularities, namely, \emph{trisectors} and \emph{wedges}.  A
trisector singularity has three hyperbolic sectors, where tensorlines
lead away from the singularity in both directions.  A wedge singularity
has a hyperbolic sector and a parabolic sector, where tensorlines
lead away in one direction and toward the singularity in the other,
possibly reduced to a single tensorline.

\subsection{Shearless-parabolic LCS}

The geodesic theory for \emph{Lagrangian coherent structures}
(\emph{LCS}) of \emph{shearless-parabolic} type, which are the focus
here, seeks such LCS as material curves with neighborhoods showing
no leading order change in along-curve-averaged \emph{Lagrangian
shear} \cite{Farazmand-etal-14}.  

The Lagrangian shear is defined as the tangential projection at
time $t$ of the advected image under the linearized flow of a unit
normal $n_0$ to a time-$t_0$ material curve at position $x_0$, i.e.,
\begin{equation}
  \sigma_{t_0}^t(x_0;e_0) := \langle\D{F}_{t_0}^t(x_0)n_0, e_t\rangle
  \equiv \frac{\langle e_0, D_{t_0}^t(x_0)e_0\rangle}{\sqrt{\langle
  e_0, C_{t_0}^t(x_0)e_0\rangle}}, 
  \label{eq:sga}
\end{equation}
where $e_0 \perp n_0$, $e_t$ is the unit tangent to the advected
curve at position $F_{t_0}^t(x_0)$, and
\begin{equation}
   D_{t_0}^t(x_0) := \smash{\frac{1}{2}}(C_{t_0}^t(x_0)\Omega -
   \Omega C_{t_0}^t(x_0))
\end{equation}
with $\Omega$ an anticlockwise rotation.  A stationary solution
$[s_1,s_2] \mapsto r(s)\in \gamma$ to the above variational
principle, namely, to
\begin{equation}
  \delta\int_{s_1}^{s_2}\sigma_{t_0}^t\left(r,\frac{r'}{|r'|}\right)\d{s}
  = 0
\end{equation}
for all admissible variations of $\gamma$, satisfies \eqref{eq:drds}
connecting CG singularities. As shown by Farazmand et al.
\cite{Farazmand-etal-14}, such admissible $\gamma$ variations are
subjected to $C_{t_0}^t(r(s_1))\allowbreak = C_{t_0}^t(r(s_2)) =
\Id$, which makes the boundary terms vanish.  

\begin{figure}[t!]
  \centering%
  \includegraphics[width=.55\textwidth]{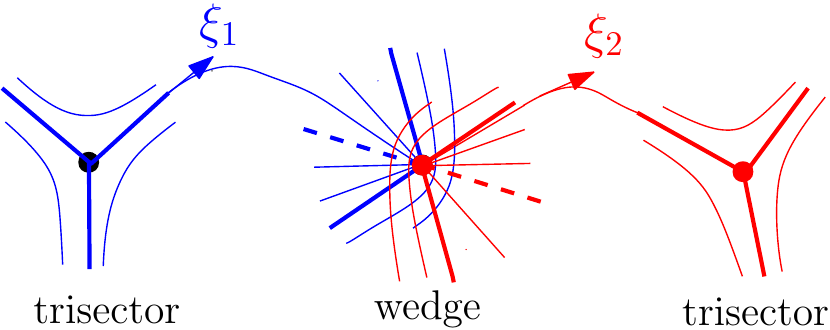}%
  \caption{Cartoon of a shearless-parabolic LCS composed of nearly
  neutral squeezing (blue) and stretching (red) tensorline segments
  of the Cauchy–Green tensor field that connect wedge and trisector
  singularities.}
  \label{fig:cartoon}%
\end{figure}

Shearless-parabolic LCS are then defined as \cite{Farazmand-etal-14}:
\begin{enumerate}
\item alternating chains of squeezeline and stretchline segments
connecting trisector and wedge singularities such that
\item the squeezing (resp., stretching) along each squeezeline
(resp., stretchline) segment is close to neutral.
\end{enumerate}
Figure \ref{fig:cartoon} shows a schematic depiction of a
shearless-parabolic LCS formed by nearly neutral squeezing (blue)
and stretching (red) tensorline segments of the CG tensor field
that connect wedge and trisector singularities.

Condition 1) above guarantees both lack of shear (indeed,
$\sigma(x_0;\xi_1) = 0 = \sigma(x_0;\xi_2)$) and structural stability.
Indeed, as a consequence of the wedge geometry, there can be no
unique connection between two wedges. On the other hand, as in the
case of heteroclinic orbits between saddles of an planar vector
field, trisector--trisector connections are structurally unstable.
As a consequence, the only types of tensorlines connecting two CG
singularities that are locally unique (i.e., there is a finite
angle between the two connecting tensorlines) and structurally
stable are trisector--wedge connections.

Condition 2) in turn guarantees parabolicity (as in the incompressible
case tangential squeezing (resp., stretching) is balanced by normal
repulsion (resp., attraction) exactly).  A cartoon of

Finally, we note that it can be shown \cite{Farazmand-etal-14} that
the shearless-parabolic LCS are null-geodesics of a pseudo-Riemannian
(Lorentzian) metric given by $g(\,\cdot\,,\,\cdot\,)(r) =
\langle\,\cdot\,,D_{t_0}^t(r)\,\cdot\,\rangle \equiv 0$, which
justifies calling the above a geodesic theory.

\subsection{Some details on implementation}

Singularities of the CG tensor points where the zero-level sets of
the scalar functions $f = C_{11} - C_{22}$ and $g = C_{12}$ intersect,
where $C_{ij}$ is the $(i,j)$th entry of $C$.  These intersections
can be found by linearly interpolating $f$ and $g$ along the edges
of a numerical grid \cite{Haller-Beron-13}. To filter out spurious
intersections produced by numerical noise, Farazmand et al.
\cite{Farazmand-etal-14} propose to consider only parts of the zero
level set of $f$ and $g$ on which $|\lambda_1\lambda_2 - 1| > 1$
holds.  In this paper we further restrict the search of singularities
to a band around the mean Eulerian axis of the Malvinas Current.

The classification of singularities may be conveniently carried out
using the geometric procedure proposed by Farazmand et al.
\cite{Farazmand-etal-14}.  This procedure exploits the distinguishing
geometric aspect of a trisector singularity with three separatrices
emanating from it.  Near the singularity, these separatrices are
close to straight lines. As a consequence, the separatrices will
be approximately perpendicular to a small circle centered at the
singularity.  With this in mind, a small circular neighbourhood of
radius $r > 0$ is selected around a singularity. With a rotating
radius vector $\mathbf r$ of length $r$, one computes the absolute
value of the cosine of the angle between $\mathbf r$ and $\xi_1$.
The singularity is classified as a trisector if $\mathbf r$ is
orthogonal to $\xi_1$ at exactly three points of the circle, and
parallel to $\xi_1$ at three other points, which mark separatrices
of the trisector.  Singularities not passing this test for trisectors
are classified as wedges.

The numerical detection of trisector--wedge connections proceeds
by tracking the separatrices leaving a trisector, and monitoring
whether they enter the attracting sector of a small circle surrounding
a wedge.

\section{Probabilistic tools}

\subsection{Transfer operator}

Assume that tracer evolution on a flow domain $X$ is governed by
an advection--diffusion process.  A tracer initially delta-concentrated
at position $x\in X$ therefore evolves passively $\mathcal T$ units of time
to a probability density $K(x,\cdot) \ge 0$.  Normalizing so that
the probability of getting somewhere from $x$ is 1, i.e., $\smash{\int_X
K(x,y)\d{y} = 1}$, $K(x,y)$ represents a stochastic kernel.  A
general initial density $f(x) \ge 0 $, $\smash{\int_X f(x) \d{x} =
1}$, evolves to
\begin{equation}
  \mathscr Pf(y) = \int_X K(x,y)f(x)\d{x},
  \label{eq:PF}
\end{equation}
where $\mathscr P$ is a linear transformation of the space of
densities $D(X)$ to itself, a Markov operator known as the
Perron--Frobenious operator or, more generally, a \emph{transfer
operator} \cite{Lasota-Mackey-94}.  Note that if $\xi_t$ is a random
tracer position distributed a time $t$ uniformly in set $A \subset
X$, i.e.,  $\xi_t \sim \smash{\frac{\mathbf 1_{A}}{\area(A)}}$ where
and $\mathbf 1_{A}(x) = 1$ for $x\in A$ and 0 otherwise, then the
probability to be found in $B \subset X$ after $\mathcal T$ units
of time is given by:
\begin{equation}
  \prob[\xi_{t+\mathcal T} \in B \mid \xi_t\in A] = \frac{\int_{B}\mathscr
  P\frac{\mathbf 1_{A}(y)}{\area(A)} \d{y}}{\int_A  \frac{\mathbf
  1_{A}(x)}{\area(A)} \d{x}} = \frac{1}{\area(A)}\int_{A}
  \int_{B} K(x,y)\, \d{y}\d{x}.
\end{equation}

\begin{figure}[t]
  \centering%
  \includegraphics[width=.5\textwidth]{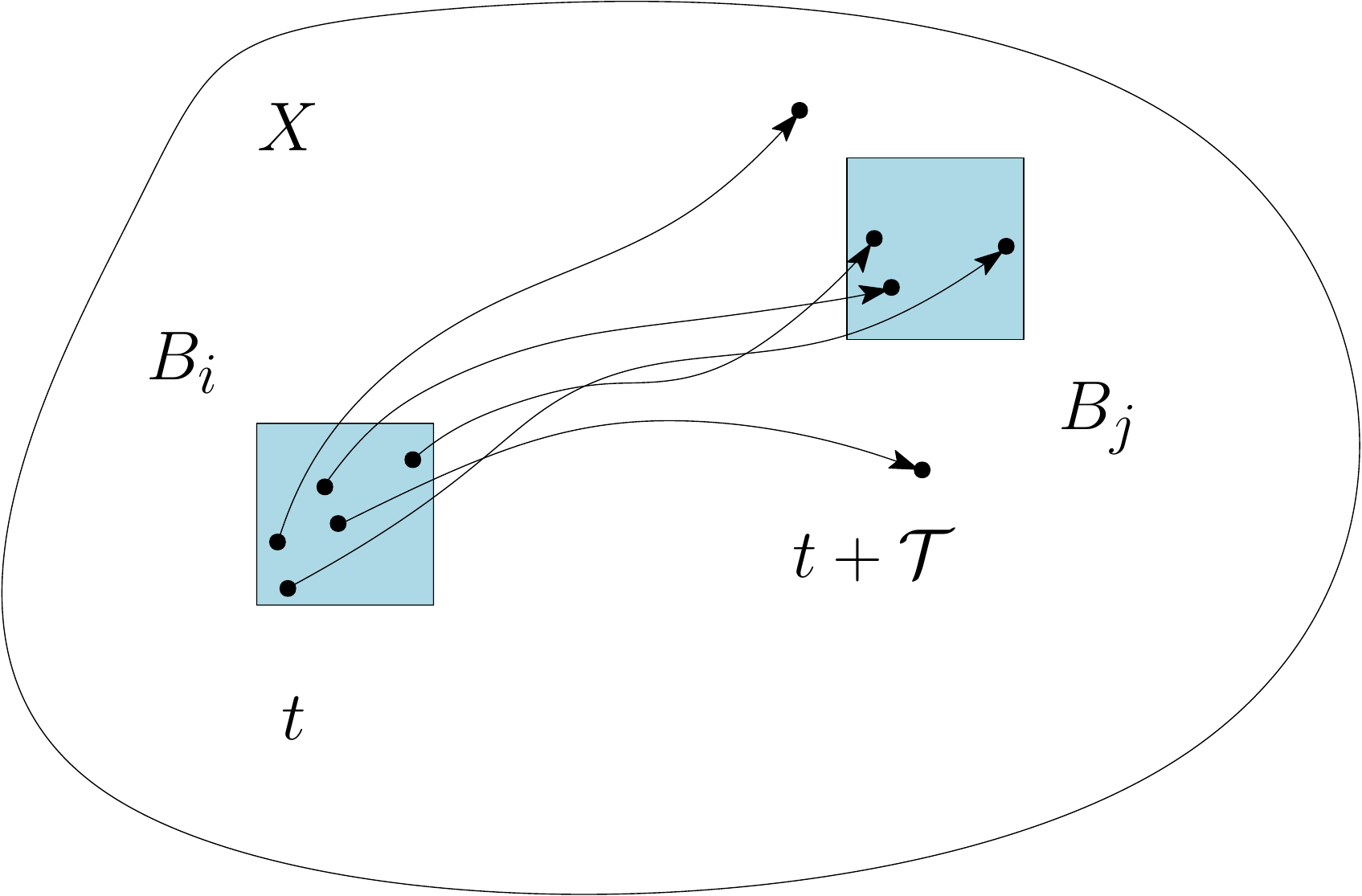}%
  \caption{The discrete action of the transfer operator $\mathscr
  P$ describing an advection--diffusion process in flow domain $X$
  over the time interval $[t,t+\mathcal T]$ is represented by an
  $N \times N$ matrix $P$ of transitional probabilities of tracer
  particles of moving between boxes $\{B_1, \dotsc, B_N\}$ covering
  $X$ from time $t$ to time $t+\mathcal T$.}
  \label{fig:P}%
\end{figure}

\subsection{Transition matrix}

Suppose that we are presented with tracer data in the form of
trajectories of individual tracer particles.  One can evaluate the
action of $\mathscr P$ on $f$ through a discretization of the
Lagrangian dynamics using Ulam's method \cite{Ulam-60}.  This
consists in partitioning the domain $X$ into a grid of $N$ connected
boxes $\{B_1, \dotsc, B_N\}$ and projecting functions in $D(X)$
onto a finite-dimensional space $V_N$ approximating $D(X)$ and
spanned by indicator functions on the grid. Specifically, $f(x)
\approx \sum_1^N f_i \smash{\frac{\mathbf 1_{B_i}(x)}{\area(B_i)}}$
where $f_i = \smash{\int_{B_i}}f(x)\d{x}$. The discrete action of
$\mathscr P$ on $V_N$ is described by an $N\times N$ matrix, called
a \emph{transition matrix}, with components given by
\begin{equation}
  P_{ij} = \int_{B_j}\mathscr P\frac{\mathbf 1_{B_i}(y)}{\area(B_i)}
  \d{y} = \prob[\xi_{t+\mathcal T} \in j \mid \xi_t\in B_i].
\end{equation} 
By considering a sufficiently large number of particles, the
components of $P$ can be estimated as \cite{Miron-etal-18a}
\begin{equation}
  P_{ij} \approx 
  \frac{\mbox{\# of particles in $B_i$
  at any $t$ that evolve to $B_j$ at
  $t+T$}}{\mbox{\# of particles in $B_i$ at
  any $t$}}.
  \label{eq:P}
\end{equation}
The matrix $P$ defines a Markov-chain representation of the dynamics,
with the components $P_{ij}$ equal to one-step conditional transition
probabilities of moving between boxes, which represent the states
of the chain. The evolution of the discrete representation of
$f(x)\in D(X)$, i.e., a vector $\mathbf{f} = (f_1\, \cdots \, f_N)$
of probability since $\smash{\sum_{i=1}^N f_i = 1}$, is calculated
under left multiplication,
\begin{equation}
  \mathbf{f}^{(k)} = \mathbf{f}P^{k},\quad k = 1, 2, \dotsc.
  \label{eq:pP}
\end{equation}

\subsection{Irreducibility, aperiodicity, and attracting sets}

Because the transition matrix $P$ is (row) stochastic, i.e.,
$\smash{\sum_{j=1}^N} P_{ij}$ for every $i$, the vector $\mathbf{1}$
of ones is a right eigenvector with eigenvalue $\lambda = 1$ (i.e.,
$P\mathbf{1} = \mathbf{1}$), which is maximal.  The associated
nonunique left eigenvector $\mathbf{p}$ is invariant (because
$\mathbf{p}P = \mathbf{p}$) and nonnegative (by the
Perron--Frobenius theorem \cite{Horn-Johnson-90}). 

If $P$ is \emph{irreducible} (i.e., $\exists\, n_{ij} < \infty :
\smash{(P^{n_{ij}})_{ij} > 0}$, meaning that all states in the
Markov chain communicate) and \emph{aperiodic} (i.e., $\exists\, i
: \gcd\{n \ge 0 : (P^n)_{ii} > 0\} = 1$, meaning that no state is
revisited cyclically), then the eigenvalue $\lambda = 1$ is simple,
the corresponding left eigenvector $\mathbf{p}$ is strictly positive,
and (scaled to a probability vector) $\smash{\mathbf{p} =
\lim_{k\uparrow\infty} \mathbf f P^k}$ for any initial probability vector
$\mathbf f$.

Suppose that $P$ is irreducible on some class of states $S \subset
\{1,\dotsc,N\}$.  We call $S$ an \emph{absorbing closed communicating
class} if $\smash{P_{ij}} = 0$ for all $i \in S$, $j\notin S$, and
$\smash{P_{ij}} > 0$ for some $i \notin S$, $j\in S$;  cf.\ Froyland
et al. \cite{Froyland-etal-14}.  The set $B_S = \smash{\bigcup_{i\in
S}} B_i\subset X$ forms an approximate \emph{time-asymptotic
forward-invariant attracting set} for trajectories starting in
$\smash{X = \bigcup_{i=1}^N B_i}$

\subsection{Communication}

In practice a Markov chain can be viewed as a directed graph with
vertices in the graph corresponding to states in the chain, and
directed arcs in the graph corresponding to one-step transitions
of positive probability.  This allows one to apply Tarjan's algorithm
\cite{Tarjan-72} to assess communication within a chain.  Specifically,
the Tarjan algorithm takes such a graph as input and produces a
partition of the graph's vertices into the graph's strongly connected
components.  A directed graph is strongly connected if there is a
path between all pairs of vertices.  A strongly connected component
of a directed graph is a maximal strongly connected subgraph and
by definition also a maximal communicating class of the underlying
Markov chain.

\subsection{Almost-invariant sets}

Revealing those regions in which trajectories tend to stay for a
long time before entering another region is key to assessing
connectivity in a flow.  Such \emph{forward time-asymptotic
almost-invariant sets} and their corresponding \emph{backward-time
basins of attraction} can be framed \cite{Froyland-etal-14} by
inspecting eigenvectors of $P$ with $\lambda \approx 1$.

The magnitude of the eigenvalues quantifies the geometric rates at
which eigenvectors decay.  Those left eigenvectors with $\lambda$
closest to 1 are the slowest to decay and thus represent the most
long-lived transient modes \cite{Froyland-97, Pikovsky-Popovych-03}.
For a given $\lambda \approx 1$, a forward time-asymptotic
almost-invariant set will be identified with the support of similarly
valued and like-sign elements in the left eigenvector.  Regions
where the magnitude of the left eigenvector is greatest are the
most dynamically disconnected and take the longest times to transit
to other almost-invariant sets.

\subsection{Lagrangian geography}

The multiple backward-time basins of attraction are identified by
boxes where the corresponding right eigenvectors take approximately
constant values (cf.\ Koltai \cite{Koltai-11} for the simpler single
basin case).  Decomposition of the ocean flow into weakly disjoint
basins of attraction for time-asymptotic almost-invariant attracting
sets using the above eigenvector method has been shown
\cite{Froyland-etal-14, Miron-etal-17, Miron-etal-18a} to form the
basis of a \emph{Lagrangian geography} of the ocean, where the
boundaries between basins are determined from the Lagrangian
circulation itself, rather than from arbitrary geographical divisions.

The number of provinces in a geography will depend on the number
of right eigenvectors considered.  A large gap in the eigenspectrum
of $P$ provides a cutoff criterion for eigenvector analysis, as
provinces extracted from eigenvectors with eigenvalues on the right
of the gap will have significantly shorter retention times than
those extracted from eigenvectors on the left of the gap.

\subsection{Retention time}

 Consider $\mathbf p_A P|_A = \lambda_A \mathbf
p_A$, where $P|_A$ is $P$ restricted to some set $A\subset X$ and
$\lambda_A$ is the largest eigenvalue of $P|_A$.  If $P$ is
irreducible, $\lambda_A < 1$ and $\mathbf p_A \ge 0$.  Assume that
a tracer trajectory starts in $A$.  If the trajectory is conditioned
on being retained in $A$, it will asymptotically distribute as
$\mathbf p_A$, where $\mathbf p_A$ has been normalized to a probability
vector.  Such a $\mathbf p_A$ is called a quasi-stationary distribution
\cite{Bremaud-99}.  The probability of the random time $\mathfrak
T$ to exit $A$ to be longer than $k\mathcal T$ is equal to the
probability of the trajectory to remain in $A$ after $k$ applications
of $P$, conditioned on being initially in $A$ and distributed as
$\mathbf p_A$.  This is thus
\begin{equation}
  \prob[\mathfrak T > k\mathcal T] = \prob[\xi_{t+k\mathcal T} \in
  A \mid \xi_t \in A] = \frac{\sum_i (\mathbf p_A
  (P|_A)^k)_i}{\sum_i (\mathbf p_A)_i} = \lambda_A^k.
\end{equation}
Now note that 
\begin{equation}
  \sum_0^\infty\prob[\mathfrak T > k\mathcal T] = \sum_0^\infty
  k \prob[\mathfrak T = k\mathcal T].
\end{equation}
Therefore, the expectation of the random exit time is
\begin{equation}
  \Exp[\mathfrak T] = \mathcal T\cdot \sum_0^\infty
  k \prob[\mathfrak T = k\mathcal T] = \mathcal
  T\cdot\sum_0^\infty \lambda_A^k = \frac{\mathcal T}{1 -
  \lambda_A},
\end{equation}
which provides an average measure of retention time in $A$.

\section{Symmetric stability}

Here we adopt a coordinate system with $\mathbb R^3\{x,y,z\}$ with
$z$ pointing upward.  Dropping out $y$ dependencies, Cho et al.
\cite{Cho-etal-93} show that the inviscid, unforced, nonhydrostatic,
Boussinesq equations on an $f$ plane form a Lie--Poisson system on
$\varphi := (\omega_y, m, b)$ with Hamiltonian (energy)
\begin{equation}
  \mathscr E[\varphi] = - \int \big(\tfrac{1}{2}\psi\omega_y + fxm +
  zb\big)\d{x}\d{z}.
\end{equation}
Here $\omega_y := \partial_z u - \partial_x w$ is the vorticity in
the $y$-direction, which in terms of the streamfuction $\psi$ reads
$\omega_y = \nabla^2\psi$, where $\nabla := (\partial_x,\partial_z)$;
$m := v + fx$, $b := - g\rho/\rho_0$ is the buoyancy, where $g$ is
gravity, and $\rho$ and $\rho_0$ density and reference (constant)
density, respectively; and integration is on a closed domain or the
whole plane with appropriate decaying conditions at infinity.
Explicitly,
\begin{eqnarray}
  \partial_t \omega_y &=& [\omega_y,\psi] + [m,fx] +
  [b,z],\\
  \partial_t m &=& [m,\psi],\\
  \partial_t b &=& [b,\psi],
\end{eqnarray}
using the Lie--Poisson bracket $\{\mathscr A, \mathscr B\} := \sum_i
\int \varphi^i [\delta \mathscr A/\delta \varphi^i, \delta \mathscr
B/\delta \varphi^i] \d{x}\d{z}$ for admissible functionals
$\mathscr A,\mathscr B[\varphi]$ where $[\,,]$ is the Jacobian
bracket. 

The Casimirs of the systems are of the form
\begin{equation}
  \mathscr C[\varphi] = \int C(m,b)\d{x}\d{z}
\end{equation}
for all $C(\cdot,\cdot)$. Their conservation is a consequence of
the material conservation of $m$ and $b$ by the flow.  Note that
the potential vorticity 
\begin{equation}
  q := (\omega_x,f+\omega_z)\cdot\nabla b \equiv [m,b],
\end{equation}
so is materially conserved. 

A steady baroclinic parallel flow in thermal wind balance $f\partial_z
V = \partial_x B$ represents an equilibrium of the system, including
$V(x,z) = S_0^2z/f$ and $B(x,z) = S_0^2x + N_0^2z$ (where $S_0^2$
and $N_0^2>0$ are arbitrary constants) as originally considered by
Hoskins \cite{Hoskins-74}.  Convexity of the conserved pseudoenergy
\begin{equation}
  \mathscr H[\delta\varphi;\Phi] := (\mathscr E + \mathscr
  C)[\omega,m,b] - (\mathscr E + \mathscr C)[0,M,B]
  \label{eq:H}
\end{equation}
guarantees stability of the basic flow under finite-amplitude
perturbations of arbitrary shape \cite{Arnold-65, Holm-etal-85}.

Using
\begin{equation}
  C(m,b) = \left(m + \frac{f}{\partial_x M \partial_z
  M}b\right)^2 + \frac{\partial_x M}{Q}b^2,
\end{equation}
where
\begin{equation}
  Q =  [M,B] = N^2(f + \partial_x V) - 
  f(\partial_z V)^2
\end{equation}
is the basic flow's potential vorticity with $N^2 := \partial_z B$
the basic flow's Brunt--V\"ais\"al\"a frequency squared, Cho et al.
\cite{Cho-etal-93} show that there exist constants $0 < c_- \leq
c_+ < \infty$ such that
\begin{equation}
  \Vert\delta \varphi\Vert^2_{t>t_0} \le \mathscr H|_{t>t_0} =
  \mathscr H|_{t=t_0} \le \frac{c_+}{c_-}\Vert\delta\varphi\Vert^2_{t=t_0}
\end{equation}
where 
\begin{equation}
  \Vert\delta \varphi\Vert^2 := \frac{1}{2}\int
  \big(|\nabla\delta\psi|^2 + c_-\delta m^2 +
  c_- N^{-2}\delta b^2\big)\d{x}\d{z},
\end{equation}
iff
\begin{equation}
  \mathrm{Ri} := \frac{N^2}{(\partial_z V)^2} > \frac{1}{1 +
  \partial_x V/f} > 0 
  \label{eq:sym-app}
\end{equation}
where Ri is the Richardson number (cf.\ also Mu et al.  \cite{Mu-etal-96}).
Conditions \eqref{eq:sym-app} guarantee convexity of \eqref{eq:H}
and thus normed (i.e., Lyapunov) stability.

Cho et al. \cite{Cho-etal-93} use a virial functional \cite{Chandrasekhar-69}
to further show that \eqref{eq:sym-app} are both sufficient and
necessary conditions for stability.  We note that, for a stably
stratified fluid ($N^2 > 0$), \eqref{eq:sym-app} reduce to $fQ >
0$, which is preserved under the flow. Thus $fQ < 0$ is both necessary
and sufficient for instability.

The necessary and sufficient condition for nonlinear stability $fQ
> 0$ clearly includes the sufficient condition for the stability
of the basic state  $V(x,z) = S_0^2z/f$ and $B(x,z) = S_0^2x +
N_0^2z$ under $y$-independent infinitesimally small normal-mode
perturbations originally derived by Hoskins \cite{Hoskins-74}.


\end{document}